\documentclass[12pt,a4paper,oneside]{book}

\usepackage[hyperindex,breaklinks]{hyperref}
\usepackage{graphicx}
\usepackage[english]{babel}
\usepackage{amsmath}  
\usepackage{amsfonts}
\usepackage{amssymb}
\usepackage{graphicx} 
\usepackage{xspace}
\usepackage{color}
\usepackage{url}
\usepackage{amsthm}
\usepackage{epsfig}
\usepackage{graphics}
\usepackage{float}

\usepackage{ulem}
\usepackage{color}
\def\bm{\boldmath}

\def\ra{\rightarrow}

\def\K0bar{\overline{K^0}}
\def\bge{\begin{equation}}
\def\ene{\end{equation}}
\def\bg{\begin{eqnarray}}
\def\en{\end{eqnarray}}
\def\nn{\nonumber}
\def\ra{\rightarrow}

\def\del{\partial}

\def\qbar{{\overline{q}}}
\def\ubar{{\overline{u}}}
\def\dbar{{\overline{d}}}
\def\sbar{{\bar{s}}}
\def\cbar{{\bar{c}}}
\def\d0bar{{\bar{D}^0}}
\def\Dbar{{\bar{D}}}

\def\bm{\boldmath}

 \def\nn{\nonumber}
\def\bm{\boldmath}

\def\del{\partial}

\def\ubar{\overline{u}}
\def\dbar{\overline{d}}
\def\sbar{\bar{s}}
\def\bbar{\overline{b}}
\def\Bbar{\overline{B}}
\def\cbar{\bar{c}}
\def\qbar{\overline{q}}

\def\Dbar{\overline{D}}
\def\Bbar{\overline{B}}

\newcommand{\be}{\begin{equation}}
\newcommand{\ee}{\end{equation}}
\newcommand{\bea}{\begin{eqnarray}}
\newcommand{\eea}{\end{eqnarray}}

\usepackage[T1]{fontenc}
\usepackage{ae}
\usepackage{amsmath}

\begin{document}


\begin{titlepage}     
  \begin{center}
    \vspace{\fill}
    
    \LARGE{Samuel Luiz Pinheiro Gon\c{c}alves Beres}
    \vspace{3.0cm}
    \par
    \LARGE{\bf\bm $B_{c}, B^{*}_{c}, B_{s}, B^{*}_{s}, D_{s}$ and $D^{*}_{s}$
    mass shift\\ in a nuclear medium}
    \par\vfill
    \Large{S\~ao Paulo, Março de 2024}
  \end{center}
\end{titlepage}


\begin{titlepage}
  \begin{center}
    \large{\textsc{Universidade Cidade de S\~{a}o Paulo} \\
           \textsc{Programa de P\'{o}s-Gradua\c{c}\~{a}o} \\ 
           \textsc{em Astrof\'{i}sica e F\'{i}sica Computacional} \\
          }
    \par\vfill
    \LARGE{Samuel Luiz Pinheiro Gon\c{c}alves Beres}
    \par\vfill
    \LARGE{\bf\bm $B_{c}, B^{*}_{c}, B_{s}, B^{*}_{s}, D_{s}$ and $D^{*}_{s}$ mass shift\\ in a
nuclear medium}
    \par\vfill
    \Large{S\~ao Paulo, Mar\c{c}o de 2024}
  \end{center}
\end{titlepage}


\pagenumbering{roman}
\setcounter{page}{2}


\begin{center}
\Large{\textsc{\bf Samuel Luiz Pinheiro Gonçalves Beres}}

\vfill

{\bf\bm $B_{c}, B^{*}_{c}, B_{s}, B^{*}_{s}, D_{s}$ and $D^{*}_{s}$ mass shift\\ in a nuclear
medium}

\vspace{3.0cm}

\begin{flushright}
\begin{minipage}{0.60\textwidth}
{Texto apresentado ao Programa de P\'{o}s-Gradua\c{c}\~ao em
Astrof\'{i}sica e F\'{i}sica \\ Computacional da Universidade Cidade de S\~{a}o
Paulo, para defesa do Mestrado, sob a orienta\c{c}\~ao do Dr.~Kazuo Tsushima.}
\end{minipage}
\end{flushright}

\vspace{3.0cm}

\vfill

S\~{a}o Paulo, Mar\c{c}o de 2024.

\end{center}

\newpage

\hspace{-10ex}
\includegraphics[scale=1.0]{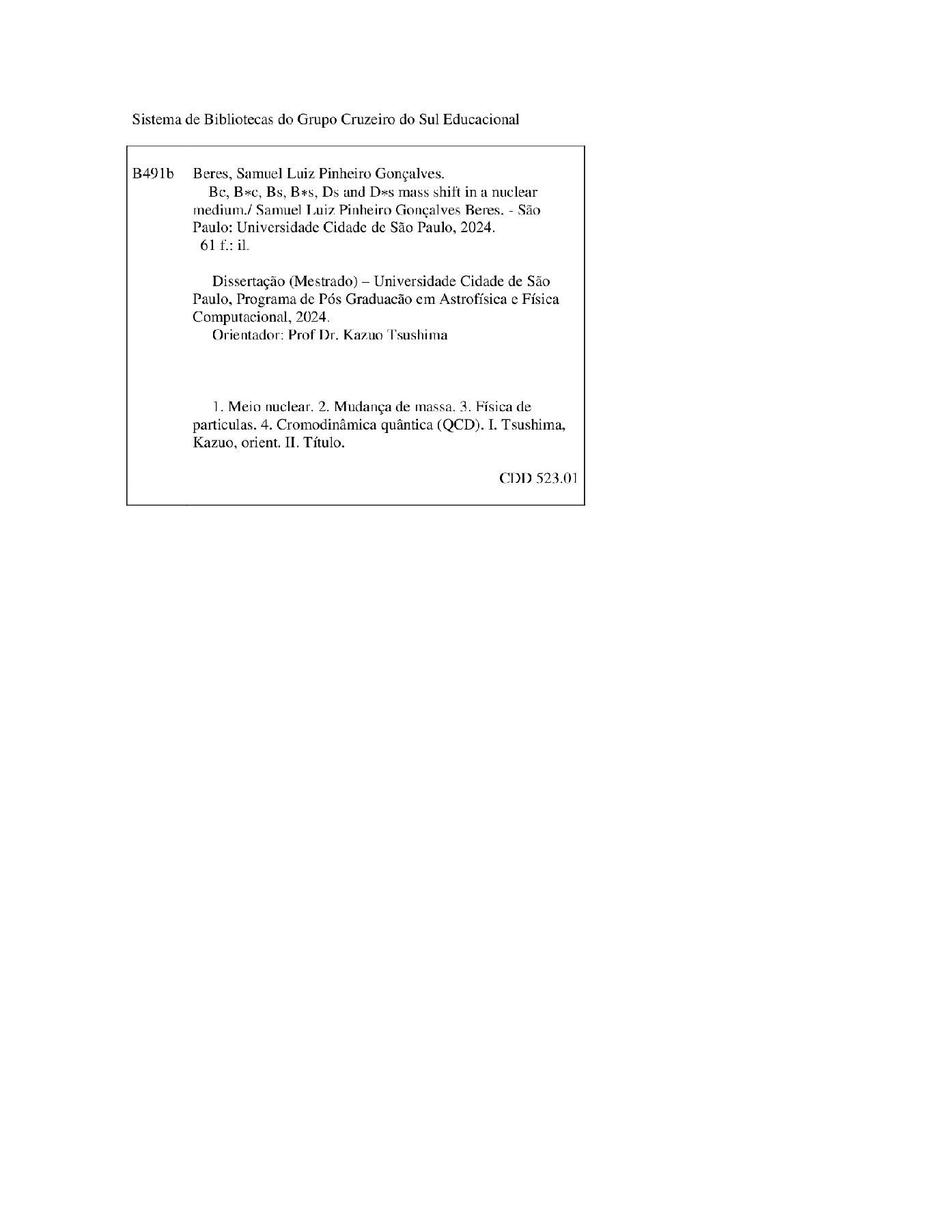}

\newpage
\thispagestyle{empty}

\begin{center}

Samuel Luiz Pinheiro Gonçalves Beres

\vspace{1.0cm}

{\Large\bf\bm
$B_{c}, B^{*}_{c}, B_{s}, B^{*}_{s}, D_{s}$ and $D^{*}_{s}$ mass shift\\ in a nuclear medium
}

\vspace{1.0cm}

\begin{flushright}

\begin{minipage}{0.5\textwidth}
{Texto apresentado ao Programa de P\'{o}s-Gradua\c{c}\~ao em
Astrof\'{i}sica e F\'{i}sica \\ Computacional da Universidade Cidade de S\~{a}o
Paulo, para defesa do Mestrado, sob a orienta\c{c}\~ao do Dr.~Kazuo Tsushima.}
\end{minipage}
\end{flushright}

\vfill

\begin{flushleft}

Aprovada em Mar\c{c}o de 2024.

\end{flushleft}

\vfill

BANCA EXAMINADORA

\vfill

\hrulefill \\Prof. Dr. Kazuo Tsushima - Orientador\\
UNIVERSIDADE CIDADE DE S\~{A}O PAULO\\

\vfill

\hrulefill \\ Prof. Dr. Jo\~{a}o Pacheco B. C. de Melo\\
UNIVERSIDADE CIDADE DE S\~{A}O PAULO\\

\vfill

\hrulefill \\Prof. Dr. Gast\~{a}o I. Krein\\
UNIVERSIDADE ESTADUAL DE S\~{A}O  PAULO\\

\vfill


S\~{a}o Paulo, Mar\c{c}o de 2024

\end{center}

\newpage



\begin{flushright}
\begin{minipage}{0.5\textwidth}

\vspace{15.0cm} 

To my Aunt and my friend Danilo for all the support
and incentives in times of storm.
To my supervisor Dr. Kazuo Tsushima, my inspiration and definition of excellence in work and research,
thank you for taught me to continuously improve.

\end{minipage}
\end{flushright}

\clearpage


\section*{Acknowledgements}

My thanks go to 
Professor Kazuo Tsushima for being so patient and for all the lessons and advice that allowed me to grow as a student
and better person, for all that I  will be forever grateful.\\
Thanks go to my senior colleague Ms. Guilherme N. Zeminiani and all my dear colleagues and professors of LFTC.
I would like to thank especially Professor Gastão I. Krein of Universidade Estadual de São Paulo (UNESP)
and Professor Fernando S. Navarra of Universidade de São Paulo (USP)
for their kindness to be the committee members of  the thesis/master defense of mine.\\
Finally thanks to Universidade Cidade de S\~ao Paulo and Dr. Gustavo A. Lanfranchi for the support and opportunity.\\
This study was financed in part by the Coordenação de Aperfeiçoamento de Pessoal de Nível Superior –
Brasil (CAPES) – Finance Code 001.

\thispagestyle{myheadings}
\clearpage



    \vspace*{\fill}
	\begin{flushright}
		\textit{``If I have seen further than others, it is by standing upon the shoulders of giants.``\\
Isaac Newton}
	\end{flushright}
\clearpage


{
\listoffigures
\addcontentsline{toc}{section}{List of Figures}
}

\thispagestyle{myheadings}


\listoftables
\addcontentsline{toc}{section}{Lista de Tabelas}

\thispagestyle{myheadings}


\tableofcontents

\thispagestyle{myheadings}


\chapter*{Resumo}
\addcontentsline{toc}{chapter}{Resumo}

Pela primeira vez, estimamos a  mudan\c{c}a de massa
dos m\'{e}sons pesados de dois sabores $B_c, B_c^*, B_s, B_s^*, D_s$ e $D_s^*$ 
em mat\'{e}ria nuclear sim\'{e}trica. As estimativas  s\~{a}o feitas avaliando
as autoenergias de um loop de ordem mais baixa. A origem dessas mudan\c{c}as de massa 
são as mudan\c{c}as nos m\'{e}sons pesados e leves do estado intermedi\'{a}rio 
dos m\'{e}sons pesados de dois sabores em mat\'{e}ria nuclear sim\'{e}trica.
Nossos resultados mostram que a magnitude da mudança de massa para
o m\'{e}son $B_c$ ($\bar{b} c$ ou $b \bar{c}$) \'{e} maior que
a muda\c{c}a massa dos m\'{e}sons $\eta_c (\bar{c} c)$ e $\eta_b (\bar{b} b)$,
diferente da expectativa de que seria o meio termo entre eles.
Enquanto, o de $B_c^*$ mostra o meio termo entre as de $J/\psi$ e $\Upsilon$.
Observamos que a excita\c{c}\~{a}o do m\'{e}son vetorial mais leve
nos loops dos m\'{e}sons pseudoescalares, faz com que a autoenergia contribua
mais para mudan\c{c}a de massa do m\'{e}son correspondente, $B_c, B_s,$ e $D_s$.
(Os principais resultados dessa tese foram apresentados no
artigo Ref.~\cite{Zeminiani:2023gqc}.)

\thispagestyle{myheadings}

\vspace{6ex}
\noindent
{\bf Palavras-chave:} meio nuclear,  mudan\c{c}a de massa, f\'{i}sica de part\'{i}culas,
cromodin\^{a}mica qu\^{a}ntica (QCD), m\'{e}sons pesados.

\chapter*{Abstract}
\addcontentsline{toc}{chapter}{Abstract}

For the first time, we estimate the in-medium mass shift of
the two-flavored heavy mesons $B_c, B_c^*, B_s, B_s^*, D_s$ and $D_s^*$ in symmetric nuclear
matter. The estimates are made by evaluating the lowest order one-loop self-energies.
The enhanced excitations of intermediate state
heavy-light quark mesons in symmetric nuclear matter are the origin of their negative mass
shift.
Our results show that the magnitude of the mass shift for the $B_c$ meson ($\bar{b} c$ or $b \bar{c}$)
is larger than
those of the $\eta_c (\bar{c} c)$ and $\eta_b (\bar{b} b)$,
different from a naive expectation that it would
be in-between of them.
While, that of the $B_c^*$ shows the in-between of the $J/\psi$ and $\Upsilon$.
We observe that the lighter vector meson excitation
in each meson self-energy gives a dominant contribution for the corresponding meson mass shift,
$B_c, B_s,$ and $D_s$.
(Main results of this thesis are presented in the article Ref.~\cite{Zeminiani:2023gqc}.)
\vspace{6ex}

\noindent
{\bf Keywords:} nuclear medium, mass-shift, particle physics,
quantum chromodynamics (QCD), heavy mesons.

\thispagestyle{myheadings}

\noindent




\pagebreak
\pagenumbering{arabic}


\chapter{Introduction}

\section{Brief Summary of Particle Physics}

The study of\,  ''what matter is made of''\, began many centuries ago in
Ancient Greece.
Many philosophers tried to answer this question, namely, to find the
arch\'{e}, the elements from where (supposedly) everything comes from.
For example, Anaximenes proposed that all forms of matter can be 
obtained by condensing or rarefying air~\cite{Halzen:1984}.
In 19th century, John Dalton, concluded that each element was composed
by atoms.
But, in 20th century with the development of quantum physics
(atomic, nuclear and elementary particle physics),
in particular during the years 1950s and 1960s, many subatomic
particles were found in collisions of particles
by high energy beams~\cite{Nambu:1985xx,Griffiths:2008zz,Kibble:2014spa,Ludlam:2003rh}.
Based on the huge amount of accumulated experimental data, human 
beings have established a theory of elementary particles and their interactions, the Standard
Model (SM)~\cite{Butterworth:2016mrp,Novaes:1999yn}

\begin{table}[htb]
\begin{center}
\begin{tabular}{|c|c|c| c c c|}
\hline
Lepton              & Mass  (MeV)                & Charge (in units of |e|)          &
$L_{e}$ & $L_{\mu}$ & $L_{\tau}$ \\
\hline
$e^{-}$            &  0.510                         &   $-1 $                 &  1       & 0

& 0 \\
$\mu^{-}$        &  105.6                         &   $-1 $                 &  0       & 1
& 0 \\
$\tau^{-}$       &   1776                          &   $-1 $                &  0       & 0
& 1 \\
\hline
$\nu_{e}$        &   $<2.2\times10^{-5}$  &  0                     & 1       & 0        & 0 \\
$\nu_{\mu}$    &   $< 0.17$                      &  0                     & 0       & 1        & 0 \\
$\nu_{\tau}$   &    $<15.5$                      &  0                     & 0       & 0        & 1 \\
\hline
\end{tabular}
\caption{Lepton properties.
In the third column entry, |e| is the absolute value of the electron charge.
$[L_e, L_\mu, L_\tau]$ is respectively the [$e$ (electron), $\mu$ (muon), $\tau$ (tauon)]
(lepton) number. The lepton number is $1$ for all the leptons, $-1$ for all the antileptons, and 0 for
non-leptons.}
\label{leptons}
\end{center}
\end{table}

\begin{table}[htb]
\begin{center}
\begin{tabular}{|c|c|c|c|}
\hline    
Flavor              & Mass (MeV)           & Charge (in units of |e|)
               &        Properties           \\
\hline
u                      &              2              & $\phantom{+}\frac{2}{3}$            & $I_{z}
= \frac{1}{2}$
\\
d                      &              5              & $-\frac{1}{3}$
& $I_{z} = -
\frac{1}{2}$ \\
s                      &            100            & $-\frac{1}{3}$
&
Strangeness = -1 \\
c                      &           1200           & $\phantom{+}\frac{2}{3} $           & Charm = +1 \\
b                      &          4200           & $-\frac{1}{3}$                                 &
Bottom =
-1 \\
t                       &        174000         & $\phantom{+}\frac{2}{3}$             & Top = +1 \\

\hline
\end{tabular}
\caption{Quark properties.
$I_z$ is the third component of the quark isospin,
(being given by the $(1/2) \tau_z$, the Pauli matrix divided by 2),
and all other quarks except u and d have $I_z$=0.
}
\label{quarks}
\end{center}
\end{table}

The elementary particles, we mean by this
the building blocks of the Standard Model
(SM), fermions particularly, leptons (see Table~\ref{leptons}) and quarks
(see Table~\ref{quarks}),
and bosons (see Table~\ref{bosons})~\cite{Menezes:2022},
where the quarks compose the ordinary matter.
Leptons interact electromagnetically and weakly, but only weakly if they are
not charged, while quarks interact via all three interactions, weak, strong and
electromagnetic.
We may consider gravitational interaction at the elementary particle level,
but the strength of the gravitational force is
approximately $10^{-36}$ times  of that of the electromagnetic force,
$10^{-29}$ times of that of the weak force,
and $10^{-38}$ times of that of the strong force 
(100 trillions trillions trillions times weaker).
Thus, we can ignore the gravitational interaction in the following, because the strength is
irrelevant for the elementary particles in~SM.

\begin{table}[htb!]
\begin{center}
\begin{tabular}{|c|c|c|c|}
\hline
Boson                               & Mass (GeV) & Spin      & Force \\
\hline
Gluon g                            & 0                  & 1         & strong \\
Photon $\gamma$           & 0                  & 1         & electromagnetic \\
W-boson $W^{\pm}$     & 80.4             & 1         & weak \\
Z-boson $Z^{0}$            & 91.2            & 1          & weak \\
\hline
Higgs-boson $H$             & 126             & 0         &gravitation \\
\hline
\end{tabular}
\caption{Boson properties.}
\label{bosons}
\end{center}
\end{table}

Hadrons, the strong-interaction particles (bound states of quarks and/or antiquarks),
have  mostly two kinds,
baryons that are composed of three quarks, mesons that are
composed of a pair of quark and antiquark.
There have also been found and established the existence for
exotic matter called tetra-quark~\cite{Yang:2020atz}, the strong-interaction bound system composed of two quarks and two antiquarks, and pentaquark~\cite{Yang:2020atz}, the strong-interaction bound system composed of four quarks and one antiquark.
All such observed particles are realized as the ''color-singlet'' states dictated
by the color gauge theory of quantum chromodynamics (QCD).

Quantum chromodynamics, 
is accepted as the theory of the strong interaction among quarks
and gluons~\cite{Olsen:2017bmm}, that is to say, the
interactions between and among hadrons.
As examples, we list the low mass baryons and mesons
in Tables~\ref{baryons} and~\ref{mesons} (see Ref.~\cite{PDG}), respectively.

\begin{table}[htb!]
\begin{center}
\begin{tabular}{|c|c|c|c|c|}
\hline
Name  &Particle label             &Quark Component   & Rest Mass (MeV) & Spin \\
\hline
Proton &p                                    & uud          & 938.3                            &
$\frac{1}{2}
$ \\
Neutron &n                                  & ddu          & 939.6                            &
$\frac{1}{2}$
\\
Lambda &$\Lambda^{0}$           & uds          & 1115.6                          &
$\frac{1}{2}$ \\
Sigma &$\Sigma^{+}$                & uus           & 1189.4                          & $\frac{1}{2}$
\\
Sigma &$\Sigma^{0}$                 & uds           & 1192.5                          &
$\frac{1}{2}$ \\
Sigma &$\Sigma^{-}$                 & dds            & 1197.3                          &
$\frac{1}{2}$ \\
Delta &$\Delta^{++}$               & uuu           & 1232                              &
$\frac{3}{2}$ \\
Delta &$\Delta^{+}$                 & uud           & 1232                              &
$\frac{3}{2}$ \\
Delta &$\Delta^{0}$                  & udd           &  1232                             &
$\frac{3}{2}$ \\
Delta &$\Delta^{-}$                   & ddd           & 1232                              &
$\frac{3}{2}$ \\
Xi &$\Xi^{0}$                             & uss            & 1315                              &
$\frac{1}
{2}$ \\
Xi &$\Xi^{-}$                              & dss            & 1321                              &
$\frac{1}
{2}$ \\
Omega &$\Omega^{-}$             & sss            & 1672                              & $\frac{3}{2}$
\\
Lambda &$\Lambda^{+}_{c}$   & udc           & 2281                              & $\frac{1}{2}$ \\
\hline
\end{tabular}
\caption{
Some typical low mass baryons and their quark components~\cite{PDG}.
}
\label{baryons}
\end{center}
\end{table}

\begin{table}[htb!]
\begin{center}
\begin{tabular}{|c|c|c|c|c|c|}
\hline
Name &Particle label           & Antiparticle label           & Quark Component
                                                     & Rest Mass (MeV)      & Spin  \\
\hline
Pion &$\pi^{+}$                     & $\pi^{-}$              & $ u\dbar$
& 139.6                        &  0 \\

Pion &$\pi^{0}$                      & $\pi^{0}$             & $\frac{u\ubar- d\dbar}{\sqrt{2}}$

& 135.0                        &  0 \\
Kaon &$ K^{+}$                     & $K^{-}$                &$ u\sbar$

& 493.7                        &  0 \\
Kaon &$K^{0}_{s}$               & $\overline{K^{0}}_{s}$       & $\frac{d\sbar+s\dbar}{\sqrt{2}}$
    &
497.6                        &  0 \\
Kaon &$K^{0}_{L}$               & $\overline{{K^{0}}}_{L}$     & $\frac{d\sbar-s\dbar}{\sqrt{2}}$
     &
497.6                        &  0 \\
Eta &$\eta^{0}$                     & $\eta^{0}$          & $\frac{u\ubar+d\dbar-2s\sbar}{\sqrt{6}}$
  &
547.9                        &  0 \\
Eta prime &$\eta'^{0}$           & $\eta'^{0}$         & $\frac{u\ubar+d\dbar+s\sbar}{\sqrt{3}}$
&
958
& 0 \\
Rho &$\rho^{+}$                   & $\rho^{-}$           & $ u\dbar$

& 770                            & 1 \\
Rho &$\rho^{0}$                    & $\rho^{0}$          & $\frac{u\ubar-d\dbar}{\sqrt{2}}$

& 770                           & 1 \\
Omega &$\omega^{0}$         & $\omega^{0}$     & $\frac{u\ubar+d\dbar}{\sqrt{2}}$

& 782                           & 1 \\
Phi &$\phi^{0}$                              & $\phi^{0}$                   & $ s\sbar $

& 1020                          & 1 \\
D-meson &$D^{+}$                               & $ D^{-}$              & $ c\dbar $

& 1869.7                     & 0 \\
D-meson &$D^{0}$                               & $\Dbar^{0}$        & $ c\ubar $

& 1864.8                       & 0 \\
D$_s$-meson &$D^{+}_{s}$                       & $ D^{-}_{s}$        & $ c\sbar $

& 1968.3                       & 0 \\
J/Psi &$J/\psi$                        & $J/\psi$                 & $c\cbar$

& 3096.9                       & 1 \\
B-meson &$B^{-}$                                & $B^{+}$               & $b\ubar$

& 5279.3                          & 0 \\
B-meson &$B^{0}$                               & $\Bbar^{0}$         & $d\bbar$

& 5279.7                          & 0 \\
B$_s$-meson &$B^{0}_{s}$                        & $\Bbar^{0}_{s}$  & $s\bbar$

& 5366.9                         & 0\\
B$_c$-meson &$B^{+}_{c}$                       & $B^{-}_{c}$          & $c\bbar$

& 6274.5                        & 0 \\
Upsilon       &$\Upsilon$                             & $\Upsilon$            & $b\bbar$
& 9460.4                       & 1\\
\hline
\end{tabular}
\caption{Table of Mesons.
}
\label{mesons}
\end{center}
\end{table}

Quarks ($q$) have six ''flavors" up ($u$), down ($d$), strange ($s$), charm ($c$), bottom
 ($b$) and top ($t$).
But they can only be deduced to exist in
nature as bound systems with two or more quarks forming a baryon ($qqq$)
or with an antiquark forming a meson ($q\qbar$).
Some exotic states were also found, like tetraquark ($qq\qbar\qbar$) and pentaquark ($qqq\qbar\qbar$)~\cite{Liu:2019zoy}.
In QCD, all of such particles form ''color singlet states'', particularly only the
color singlet states can be observed.
Equivalently, since all individual quarks have ''a color quantum number'',
no isolated quarks can be, or have been found, and this fact is called
''confinement'' in  QCD of strong interaction in~SM.

Although baryons and mesons are both hadrons, baryons follow the
Fermi-Dirac statistics (fermions), and have half-integer
spins ($\frac{1}{2}, \frac{3}{2}$, ...), while the mesons follow
the Bose-Einstein statics, and have zero or integer spins (0,1,2...).

To explain the existence of baryons, for example $\Delta^{++}$~\cite{Anderson:1952nw}
in nature as experimentally observed without contradicting the Pauli exclusion
principle,
as well as to explain the $\pi^0 \to \gamma \gamma$
decay rate correctly~\cite{Fayyazuddin:2011},
the "color charge"  was introduced  (one of the basics of QCD).

Spin and Parity can conveniently classify mesons:
vector meson (total spin = 1 and odd parity),
pseudovector (axial-vector) meson (total spin = 1 and even parity),
scalar meson (total spin = 0 and even parity), pseudoscalar meson (total 
spin = 0 and odd parity).
Also other higher spin mesons exist (spin 2) and have been found,
such as tensor mesons, etc.
While for baryons, although such easier classification does not exist,
the nucleon resonances have been found up to spin 13/2, and those of the $\Delta$
up to spin 15/2. (See Ref.~\cite{PDG}.)

\section{Motivations}

In particle and nuclear physics, nuclear matter is a kind of
idealistically considered infinitely large system, 
that consists of many nucleons (usually zero-temperature is assumed).
Specifically, it may be regarded as (partly) correspond to the environment 
inside the central region of a heavy nucleus where nucleons (protons and neutrons) interact each other through the
strong force.
In particular, symmetric nuclear matter is a system of uniform, isospin and
spin saturated infinitely large system, and (usually) zero temperature
with equal number of protons and neutrons.

Inside atomic nuclei, protons and neutrons are bound together by the 
strong force. At low energies and finite baryon densities,
the nuclear medium can significantly affect the properties of nucleons
, e.g., modifying their effective masses,
interactions, and other properties
compared to those in free-space~\cite{Tsushima:2002an}.
Thus, generally hadron properties are
 also expected be modified by the effects of surrounding
nuclear medium  when they are immersed in
the nuclear medium ~\cite{Brooks:2011sa,Hayano:2008vn,Ichikawa:2018woh}.

During the last decades, many scientists have tried to study the
effects of nuclear medium using various different approaches.
A lot of experimenters use
large accelerators for colliding particles with extremely high velocities/energies
to investigate the structure of hadrons, and
the production of short-lived forms of matter.

At modern major experimental facilities in the world,
such as JLab (Thomas Jefferson Accelerator Facilities, USA),
RHIC (Relativistic Heavy Ion Collider, USA),
Fermi Lab (Fermi National Accelerator Laboratory, USA),
BNL (Brookhaven National Laboratory, USA),
J-PARC (Japan Proton Accelerator Research Complex),
SPring-8 (Super Photon Ring - 8 GeV),
DESY (Deutsches Elektronen-Synchrotron, Germany),
FAIR (Facility for Antiproton and Ion Research, Germany),
COSY (Cooler Synchrotron),
CERN (Conseil europ\'en pour la recherche nucl\'{e}aire, France),
and BES III (Beijing Spectrometer III, China),
huge number of experiments are going on and have been planned
to study the properties of hadrons~\cite{Bertulani:2015ana,Ludlam:2003rh}
in free space as well as in a nuclear medium.
 
As mentioned before, each meson is a particle composed
of a bound quark-antiquark pair~\cite{Thomas:2010nk}.
When mesons are placed in a strongly interacting medium,
such as dense nuclear environment,
properties of mesons are expected to be modified due to
the interactions with the surrounding particles (nucleons = protons and neutrons)
and generated fields in the medium~\cite{Metag:2017ixh}.

The dissociation~\cite{Hoelck:2017dby} and regeneration of mesons can occur
in extreme conditions, such as in a quark-gluon plasma (QGP) formed
in high-energy heavy-ion collisions~\cite{Harris:1996zx}.
In this QGP phase, the temperature and density are
so high that quarks and gluons are no longer confined within 
hadrons~\cite{Pasechnik:2016wkt}, and the mesons can dissociate into their
constituents, quarks and antiquarks~\cite{Braga:2018zlu}.
Meson regeneration can occur in QGP, i.e., the process 
by which mesons are recreated within the QGP medium.
However, recall that, according to QCD, we cannot directly observe
the QGP medium or isolated quarks.

The width of a meson, which is associated with the possible meson decay and other hadron creation
channels, can be modified from its free space 
when it is in medium, because of the interactions~\cite{Eletsky:1998dj}.
Often it is referred to as width broadening,  and it can provide
insights into the scattering and absorption processes involving mesons.

In a nuclear medium, mesons can couple to different
mesons and baryons with different spins and isospins.
The interactions between meson and the surrounding nuclear medium
can generate very interesting phenomena, in particular, a formation of
meson-nucleus bound states.

A meson-nucleus bound state is an exotic state which
a meson is bound in a nucleus by the attractive strong
interaction (potential) between the meson and the nucleus~\cite{Metag:2017yuh}.

One of the most prominent such in-medium modifications may be the change in
the ''effective'' mass of mesons. 
In free space, mesons have well-defined observed physical masses.
However, in a nuclear medium, the interactions between mesons and nucleons
can lead to shift in their Lorentz scalar ''masses'',
or can add Lorentz scalar potentials~\cite{Morones-Ibarra:2010sac}.
This phenomena that we referred to as ''mass shift'' of mesons in a medium,
is  consequences of strong interactions  of mesons and surrounding nuclear medium.
This is probably one of the most interesting topics in nuclear and hadron physics.
This ''mass shift'', in particular, of two-flavored heavy mesons, is
the main subject of this thesis.

With many attempts to find the evidence and understand
the origin of hadron masses and mass shift, it is clear that one has to
study and understand well their properties/masses in free space.
Hadron property change in medium has been extensively studied through various theoretical and
experimental approaches~\cite{Chhabra:2020dsr,deMelo:2018hfw,Hilger:2012zhi,Metag:2007hq,Maciel:2012,Arifi:2023jfe,Strauch:2010nm,ATLAS:2022exb}.

In the presence of nuclear medium, mesons 
are expected to change their properties due to the strong and many-body
interactions among the mesons and nucleons (hadrons).
These interactions would lead to shift in energies (Lorentz vector) and Lorentz scalar
properties, which, according to the mass-energy equivalence principle,
result in ''changes'' in meson effective masses (mass shift).

This meson mass shift can be attributed to
the partial restoration of chiral
symmetry---restoration of spontaneous and/or dynamical
chiral symmetry, since the negative mass shift is associated with the reduction in the magnitude of (light quark) quark condensate (the order parameter of chiral symmetry).

In the presence of a nuclear medium, this symmetry breaking pattern
can be modified and would be realized as mass
modifications of mesons~\cite{Lee:2021,Lee:2023ofg} (negative mass shift).

In table~\ref{MesMasTab} we present the density dependence of
some meson effective masses in MeV calculated in
the quark-meson coupling (QMC) model~\cite{Tsushima:2020gun}.
We will explain later in detail how the ''effective meson masses''
for $B_c, B_c^*, B_s, B_s^*, D_s$ and $D_s^*$ in
symmetric nuclear matter are calculated in Table~\ref{MesMasTab}.
\begin{table}[htb!]
\caption{
Density dependence of meson (effective) masses in MeV calculate in the QMC
model with $m_{u,d}=5$ MeV, $m_s=250$ MeV, $m_c=1270$ MeV and
$m_b=4200$ MeV. (See Ref.~\cite{Tsushima:2020gun}.)
($\rho_0 = 0.15$ fm$^{-3}$ below.)
Mass is independent of the charge states when we ignore the isospin symmetry breaking
like below. Note that, for the $m_{B^{*}_{c}}$ value we take the average value in Ref.~\cite{Martin-Gonzalez:2022qwd}.
\label{MesMasTab}
}
\begin{center} 
\begin{tabular}{c|c|c|c|c}
\hline
\hline
           &$\rho_B=0$ &$\rho_B=\rho_0$ &$\rho_B=2\rho_0$ &$\rho_B=3\rho_0$\\
\hline
\hline
$m_K$       &493.7  &430.5  &393.6  &369.0  \\
$m_{K^*}$   &893.9  &831.9  &797.2  &775.0  \\
$m_D$       &1867.2 &1805.2 &1770.6 &1748.4 \\
$m_{D^*}$   &2008.6 &1946.9 &1912.9 &1891.2 \\
$m_B$       &5279.3 &5218.2 &5185.1 &5164.4 \\
$m_{B^*}$   &5324.7 &5263.7 &5230.7 &5210.2 \\
$m_{B_{c}}$   &6274.5  &  &  &  \\
$m_{B^{*}_{c}}$   &6333.0  &  &  &  \\
$m_{B_{s}}$   &5366.9  &  &  &  \\
$m_{B^{*}_{s}}$   &5415.4  &  &  &  \\
$m_{D_{s}}$ &1968.4  &  &  &  \\
$m_{D^{*}_{s}}$ &2112.2  &  &  &  \\
\hline   
\hline
\end{tabular}
\end{center}
\end{table}

The study of hadron properties in a nuclear medium is of fundamental
to nuclear physics, which deals with the
structure and behavior of atomic nuclei.
This is crucial for understanding the strong force or QCD,
which is responsible for binding the protons and
neutrons together in the nucleus.

Thus, understanding the properties of nuclear matter is naturally important
for various fields, even such as astrophysics and cosmology,
since it plays a crucial role in the early universe,
in tight connection with the structure of neutron star and compact star
structure.

\section{Why $B_c$ and $B^{*}_c$ ?}

Among many particles found so far within the category of standard model (SM),
$B_c$ and $B^{*}_c$ mesons can play special roles for studying
the interactions between the heavy quarks and light quarks ($u$ and $d$),
especially the two-flavored heavy meson ($B_c$ or $B^{*}_c$) interactions with the nuclear medium.

$B_c$ and $B^{*}_c$ mesons are composed by two heavy flavors of quarks, bottom ($b$) and
charm {($\bar{c}$), i.e., $b\bar{c}$ (or $\bar{b}c$)}, making them invaluable in 
probing dynamics of the strong force within quantum chromodynamics
(QCD) in the nuclear medium,
since nuclear medium itself does not contain
heavy quarks $b$ ($\bar{b}$) or $c$ ($\bar{c}$).

This fact, based on the OZI-rule~\cite{Okubo:1963fa,Zweig:1964jf,Iizuka:1966fk},
suppresses that interactions of $B_c$ and $B^{*}_c$ mesons with the light-flavored meson
and light baryon exchanges in the lowest order,
and the interactions should be dominated by the gluons.
Thus, we have good chances to explore the roles of gluons for the interactions
of these mesons and the nuclear medium.

This will provide us with the opportunity to compare the strengths of meson-(nuclear matter)
interactions among quarkonia and two-heavy flavored mesons.
Another interest is that, whether or not the naive expectation that the
(bottom-charmed meson)-(nuclear matter) interaction strength
is in-between those of the (charmonia-nuclear matter) and (bottomonia-nuclear matter).

The $B_c$ meson (spin-0 pseudoscalar),
first theoretically studied in the early 1980s,
and subsequently observed experimentally in
1998 at Fermi National Accelerator Laboratory (Fermi Lab)~{\cite{Abe1998ObservationOB}},
to have a relatively longer lifetime compared to other mesons.
This, along with its unique quark composition, makes the $B_c$ meson an ideal
candidate for investigating its properties, in particular the interaction
with the nuclear medium. 
$B^{*}_c$ meson (spin-1 vector), an excited state of the $B_c$ meson,
can offer equally excellent opportunities for exploration.
Specifically, the $B_c$ and $B^{*}_c$ mesons serve
as sensitive probes for exploring CP 
violation, flavor physics, and the search for rare decays.
Due to the fact that the $B_{c}$ meson is composed of two heavy quarks with two
different flavors $b$ and $c$, it is naively expected to posses intermediate properties
in both mass and size of quarkonia composed
of $(b\overline{b})$ and $(c\overline{c})$.
Besides that, its ground state can not annihilate into gluons or photons,
what provides a special role compared to the quarkonium to examine heavy quark properties.

Furthermore, we focus on the mesons $B_s$ and $B_s^*$ (composed of
$b\sbar$ and $\bbar s$) as well $D_s$ and $D_s^*$ (composed of $c\sbar$ and $\cbar s$).

Since such two different non-light flavored meson
ground states cannot annihilate into gluons or photons,
they also play special roles compared to the quarkonia
($c\bar{c}$ and $b\bar{b}$ mesons).

By the project of the present thesis, therefore,
we can deepen our understanding on the strong
interactions focusing on the heavy quarks.
Moreover, although indirectly, we can study the
effect of partial restoration of chiral symmetry in a nuclear medium,
by studying the self-energies of such two-heavy-quark-flavored mesons
interacting with the nuclear medium
via the excitations of mesons with heavy-light quarks in the intermediate
states~\cite{Krein:2010vp,CobosMartinez:2021bia,Cobos-Martinez:2020ynh,Zeminiani:2021xvw,
Zeminiani:2021vaq}.

Recently, studies on the properties $B_c$ and $B_c^*$ (indeed, the two-flavored
heavy) mesons have become one of the very active
fields~\cite{Martin-Gonzalez:2022qwd,Gershtein:1994jw,Zeminiani:2020aho,Liu:2022bdq,Karyasov:2016hfm,CMS:2022sxl,
LHCb:2014mvo,LHCb:2015pbr,Charles:2020dfl,Artuso:2009jw,Borysova:2022nyr,
Arnaldi:2023zlh,Li:2023wgq,Alonso-Alvarez:2023mgc,Kim:2023htt,Penalva:2023snz,Lodhi:2011zz,Barone:2023tbl}.
Currently, experimental efforts to confirm the existence of $B_c^*$ meson is probably one of the
most important issues.

\chapter{Meson Mass-Shift Calculation \label{massshift}}

We interpret the mass shift as the in-medium Lorentz-scalar potential for the meson,
and it is computed by the difference of the in-medium meson mass ($m^*$) and the free space meson mass ($m$):

\begin{equation}
V =  m^{*} - m.
\end{equation}

The mass shift of the mesons mentioned above
are estimated by calculating the self-energies with the excitations of intermediate state mesons
with light quarks $u$ or $d$.

We employ an SU(5) effective Lagrangian (density) with
minimal substitutions to get the interaction Lagrangians
for calculating the self-energies of two-flavored heavy mesons
in free space as well as in symmetric nuclear matter.
The necessary inputs for calculating the in-medium meson self-energies are
the effective masses (Lorentz-scalar) of the intermediate state mesons (the Lorentz-vector
potentials cancel between the excited mesons appearing in the self-energy loop, and no need in the
present study).
The in-medium masses of the excited mesons appearing in the self-energy processes are calculated by
the quark-meson coupling (QMC) model invented by Guichon~\cite{Guichon:1987jp}.

\section{The Quark-Meson Coupling Model}

To get an insight in the internal structure change of
hadrons in a nuclear medium, the QMC model plays
a crucial role \cite{Saito:2005rv,Tsushima:1997df,Tsushima:2019wmq,Krein:2013rha,Tsushima:2011kh,Cobos-Martinez:2017vtr,Cobos-Martinez:2017woo,Cobos-Martinez:2017onm,Cobos-Martinez:2017fch,Tsushima:2011fg,Krein:2017usp,Stone:2016qmi,Guichon:2018uew}, similarly, for this project, the in-medium masses of mesons
$B, D$ and $K$ ($B^*$, $D^*$ and $ K^*$)
were estimated using this model.

In this model, the relativistically moving, confined light quarks $u$ and $d$
(not the heavier $s, c, b$ quarks)
in the nucleon bag, interact self-consistently with the $\sigma$, $\omega$ and $\rho$
mean fields generated by the light quarks in the other nucleons including itself.
The self-consistent response of the bound light quarks to the mean fields
$\sigma$ and $\omega$ (also isovector filed $\rho$ in isospin asymmetric nuclear matter)
leads to a new saturation mechanism for nuclear matter. Thus,
in a nuclear medium, the internal structure of the nucleon is modified
by the (light quark)-($\sigma$ scalar meson) coupling, while the (light quark)-($\omega$ vector meson) coupling shifts the energy.
The Dirac equations for the light quarks $(q = u, d)$
and heavier quarks $Q$ (see below after equations) are given by
\begin{eqnarray*}
& &\left[ i \gamma \cdot \partial_x -
(m_q - V^q_\sigma)
\mp \gamma^0
\left( V^q_\omega +
\frac{1}{2} V^q_\rho
\right) \right]
\left( \begin{array}{c} \psi_u(x)  \\
\psi_{\bar{u}}(x) \\ \end{array} \right) = 0,
\\
& &\left[ i \gamma \cdot \partial_x -
(m_q - V^q_\sigma)
\mp \gamma^0
\left( V^q_\omega -
\frac{1}{2} V^q_\rho
\right) \right]
\left( \begin{array}{c} \psi_d(x)  \\
\psi_{\bar{d}}(x) \\ \end{array} \right) = 0,
\\
&&\left[i\gamma \cdot \partial_{x} - m_{Q}\right]\psi_{Q, \overline{Q}}\left(x\right) = 0,
\end{eqnarray*}
where $Q=s,c$ or $b$ hereafter (and the same way, $\overline{q}=\ubar$, or $\dbar$ and
$\overline{Q}= \sbar, \cbar $ or $\bbar$).

Note that only the light quarks $u$ and $d$, feel the mean field potentials
$V^q_\sigma, V^q_\omega$ and $V^q_\rho$, where $V^q_\sigma \equiv g^q_\sigma \sigma$,
$V^q_\omega \equiv g^q_\omega \omega$, $V^q_\rho \equiv g^q_\rho b$,
with $(g^q_\sigma,g^q_\omega,g^q_\rho)$ is the corresponding
($q$-$\sigma$,$q$-$\omega$,$q$-$\rho$) coupling constant.  
In other words,  in the QMC model, the fields that directly
couple to the light quarks are the Lorentz-scalar-isoscalar, Lorentz-vector-isoscalar
and Lorentz-vector-isovector fields.

The eigenenergies for the quarks and antiquarks in a hadron
$h~( = B, B^*, D, D^*, K, \rm or\, K^*)$ in units of 
the in-medium bag radius of hadron $h$, $1/R^{*}_{h}$~\footnote{We indicate the in-medium quantity
with an asterisk '*', except for indicating the vector mesons, $B^*, D^*, K^*, B_c^*, B_s^*$ and
$D_s^*$, which may be clear.}
are given by
\begin{eqnarray}
&&\begin{pmatrix}
        \epsilon_{u}\\
        \epsilon_{\overline{u}}
       \end{pmatrix} = \Omega^{*}_{q} \pm R^{*}_{h} 
\left(V^{q}_{\omega} + \frac{1}{2}V^{q}_{\rho}\right),\\
&&\begin{pmatrix}
        \epsilon_{d}\\
        \epsilon_{\overline{d}}
       \end{pmatrix} = \Omega^{*}_{q} \pm R^{*}_{h} 
\left(V^{q}_{\omega} - \frac{1}{2}V^{q}_{\rho}\right),\\
&&\epsilon_{s,c,b} = \epsilon_{\overline{s},\overline{c},\overline{b}} =
\Omega^*_{s,c,b}.
\end{eqnarray}

While the mass of a hadron $h$ in symmetric nuclear matter $m^*_h$ is calculated by,
\begin{equation}
m^{*}_{h} = \sum_{j=q,\overline{q},Q,\overline{Q}} 
\frac{n_{j}\Omega^{*}_{j}- Z_{h}}{R^{*}_{h}} + \frac{4}{3} \pi R^{*3}_{h}B_{p},
\qquad
\left. \frac{d m^{*}_{h}}{d R_{h}}\right|_{R_{h} = R^{*}_{h}} = 0.
\end{equation}
where,
$\Omega^*_q = \Omega^*_{\overline{q}} = [x_q^{2} + ( R^*_{h} m^*_{q})^2]^{1/2}$, 
$m^{*}_{q} = m_{q} - g^{q}_{\sigma}\sigma$,
$\Omega^{*}_{Q} = \Omega^{*}_{\overline{Q}} = \left[x^{2}_{Q} + \left(R^{*}_{h} 
m_{Q}\right)^{2}\right]^{\frac{1}{2}}$,
with $x^{*}_{q,Q}$ being the lowest-mode bag eigenfrequencies, 
$B_{p}$ is the bag constant (assumed to be independent of density),
$n_{q,Q}[n_{\overline{q}\overline{Q}}]$ are the lowest-mode valence quark numbers of each quark
flavor in hadron $h$,
$Z_{h}$ parametrizes the sum of the center of mass and gluon
fluctuation effects (assumed to be
independent of density).
The current quark mass values used are
$(m_q, m_s, m_c, m_b) = (5, 250, 1270, 4200)$ MeV.
(See Ref.~\cite{Tsushima:2020gun} for the other values used,
$(m_q, m_s, m_c, m_b) = (5, 93, 1270, 4180)$ MeV.)
The free space nucleon bag radius is chosen to be
$R_{N}$ = 0.8 fm, and the light quark-meson coupling constants,
$g^{q}_{\sigma}$, $g^{q}_{\omega}$ and $g^{q}_{\rho}$, are determined by
the fit to the binding energy per nucleon of 15.7 MeV
at the saturation density ($\rho_0 = 0.15$ fm$^{-3}$) of
symmetric nuclear matter, and the bulk symmetry energy 
(35 MeV)~\cite{Guichon:1987jp,Saito:2005rv,Tsushima:1997df}.
Bag constant $B_{p} = (170 {\rm\, MeV})^{4}$ is determined by the free nucleon mass $m_{N} = 939$
MeV which is the standard input values in the QMC model.

\begin{figure}[htb!]
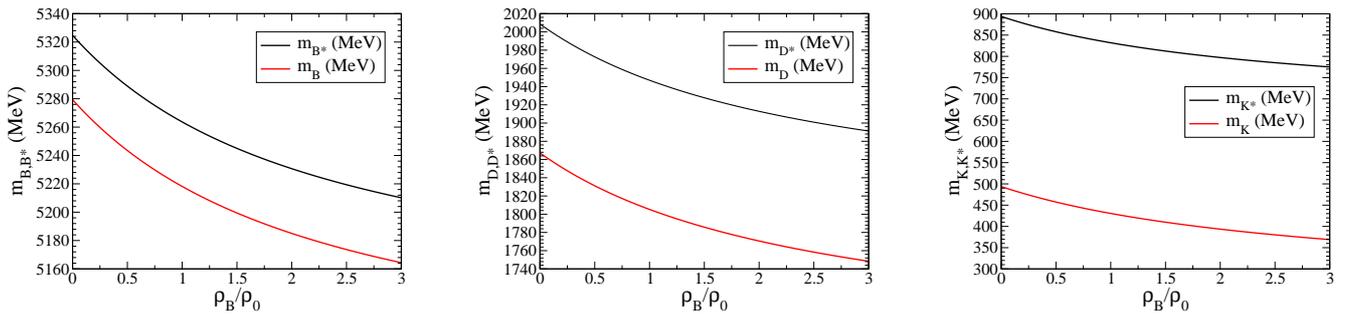

\begin{center}
\vspace{3ex}
\hspace{-8.8ex}
\includegraphics[scale=0.23]{meson_BBs_mass}\hspace{4.2ex}
\includegraphics[scale=0.23]{meson_DDs_mass}\hspace{4.4ex}
\includegraphics[scale=0.23]{meson_KKs_mass}
\caption{$B$ and $B^{*}$ (left panel), $D$ and $D^{*}$ (middle panel) and $K$ and $K^{*}$ (right
panel) meson Lorentz-scalar effective masses in symmetric nuclear matter versus baryon density
$(\rho_B / \rho_0)$, calculated by the QMC model.
\label{BDKmass}}
\end{center}
\end{figure}

In Fig.~\ref{BDKmass} it is presented the in-medium Lorentz-scalar
effective masses of mesons $B,B^{*},D,D^{*}K$ and $K^{*}$.

The QMC model predicts mass shift ($\Delta m =m^{*}-m$) 
for these meson in symmetric nuclear matter at $\rho_0$ ($\Delta m_{B}, \Delta m_{B^{*}}, \Delta m_{{D}}, \Delta m_{D^{*}}, \Delta m_{K}, \Delta m_{K^{*}}) \\
= (-61.13,- 61.05, -61.97, -61.66, -63.2, -61.97)$ MeV where $\rho_0 = 0.15$ fm$^{-3}$.
We use these density dependence of the in-medium masses to estimate the self-energies of
$B_{c}, B^{*}_{c}, B_{s}, B^{*}_{s},
 D_{s}$ and $D^{*}_{s}$ .

\section{Meson Self-Energy \label{self-energy}}

Let us analyse the lowest one-loop order self-energy diagram of particle $A$
depicted below.
\begin{figure}[htb!]
\begin{center}
\hspace{16ex}
\includegraphics[scale=0.8]{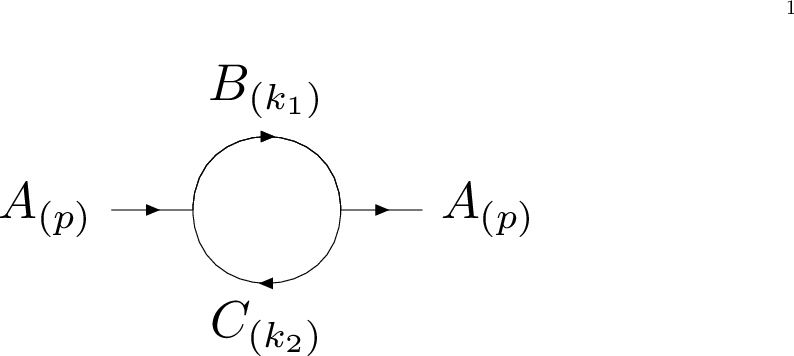}
\caption{The lowest one-loop order self-energy of particle $A$.
From the left- to right-direction we have an
incoming particle A carrying a
momentum (p)  which will reach the vertex and
dissociate into two virtual particles denoted by B and C 
carrying a respective momentum $k_1$ and $k_2$ and finally recombine into A again.
\label{selfE}}
\end{center}
\end{figure}

\begin{figure}[htb!]
\begin{center}
\includegraphics[scale=0.25]{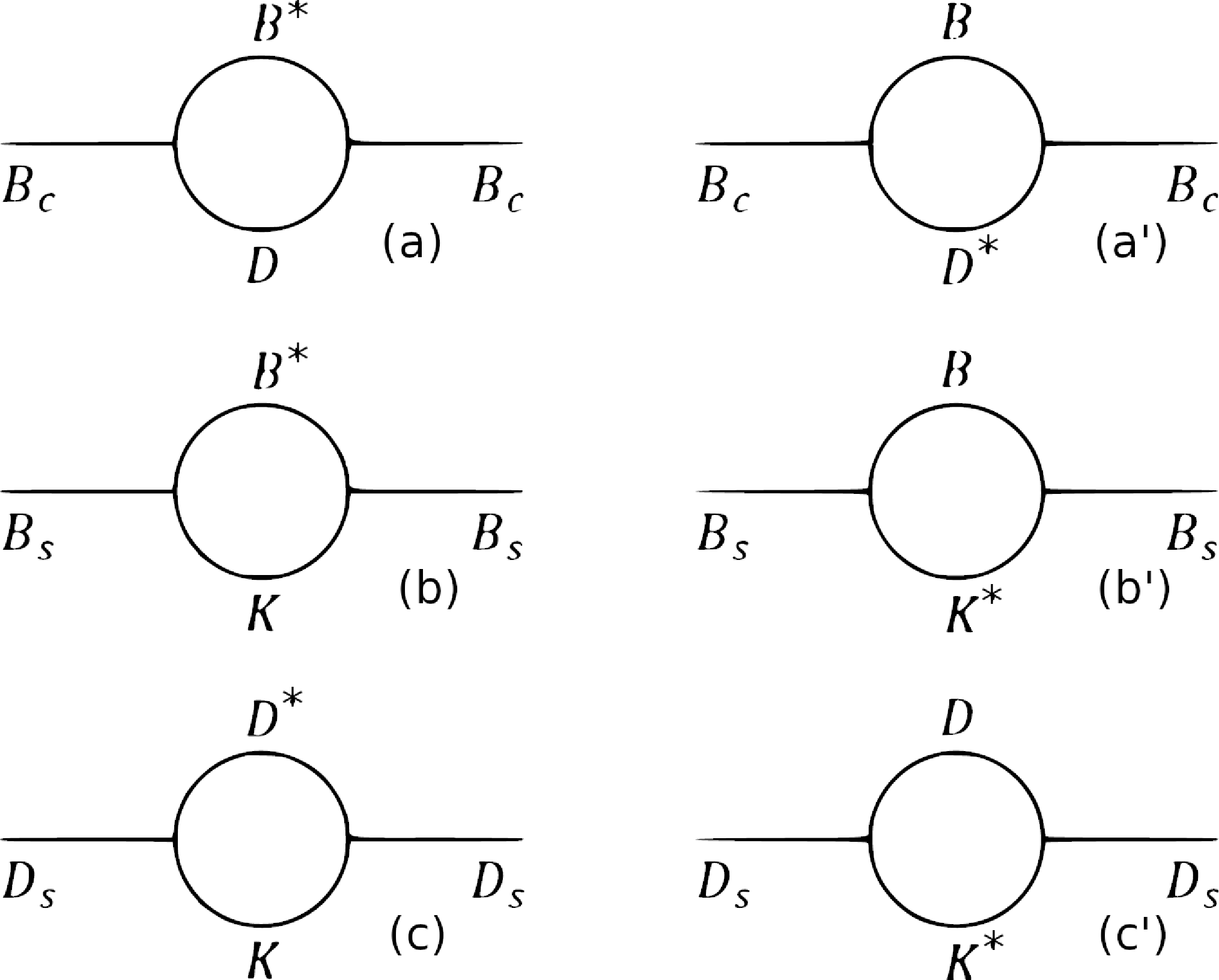}
\caption{Pseudoscalar meson self-energies
in two lighter meson excitations
in the one-loop order, for $B_c$ (top panel), $B_s$ (middle panel), and
$D_s$ (bottom panel).
\label{psmeson}}
\end{center}
\end{figure}

Based on the self-energy process shown in Fig.~\ref{selfE},
we are now in a position to explain specific cases
relevant for our studies as follows:
In Fig.~\ref{psmeson} we show the self-energy loop graphs for mesons $B_{c}$ (top),
$B_{s}$ (middle) and $D_{s}$ (bottom).
Our particular interest is at the left panel for each meson in
Fig~\ref{psmeson}, because each of them presents a lighter total mass excitation
in the corresponding meson self-energies, and thus we may naively expect
that the left graph may be more dominant than the right graph in each horizontal panel.
(However, the final results will show that this is not the case.)

We explicitly give the total mass values of the intermediate states corresponding to the
graphs shown in Fig.~\ref{psmeson}:
\bea
(B^{*},D) \rightarrow m_{B^{*}} + m_D = 7191.9\,\,{\rm MeV},
&(B,D^{*}) \rightarrow  m_B + m_{D^{*}} = 7287.9\,\,{\rm MeV},
\label{tmass1}\\
(B^{*},K) \rightarrow m_{B^{*}} + m_K = 5818.4\,\,{\rm MeV},
&(B,K^{*}) \rightarrow  m_B + m_{K^{*}} = 6173.3\,\,{\rm MeV},
\label{tmas2}\\
(D^{*},K) \rightarrow m_{D^{*}} + m_K = 2502.3\,\,{\rm MeV},
&(D,K^{*}) \rightarrow m_D + m_{K^{*}} = 2761.1\,\,{\rm MeV}.
\label{tmass3}
\eea

By the above comparison, we note that from Eq.~(\ref{tmass1}),
the $(B^*D)$ and $(BD^*)$ cases, the total mass
values are rather close, and some structure, Lorentz structure, pole position etc.
may reverse our naive expectation. (In fact, it is the case as will be shown.)

\begin{figure}[htb!]
\begin{center}
\includegraphics[scale=0.32]{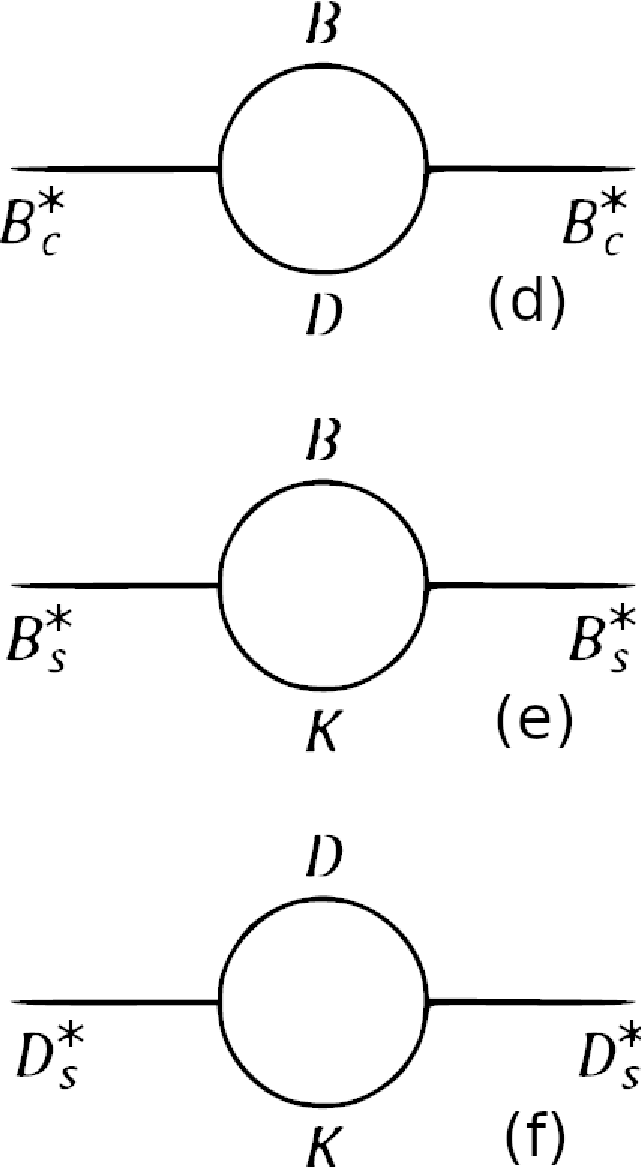}
\caption{Vector meson self-energies for
$B_{c}^{*}$ (top panel), $B_{s}^{*}$ (middle panel)
and $D_{s}^{*}$ (bottom panel) in the lowest order.
\label{vmeson}}
\end{center}
\end{figure}

Similarly, we consider the self-energy graphs for the vector mesons 
$B_{c}^{*}$, $B_{s}^{*}$ and $D_{s}^{*}$
depicted in Fig.~\ref{vmeson}

In the following we focus on the $B_c$ and $B_c^*$ mesons,
respectively the top panel in Fig.~\ref{psmeson},
and the top panel in Fig~\ref{vmeson}.
For the other meson self-energies in Figs.~\ref{psmeson} and~\ref{vmeson},
we give the detailed results in  later sections.

Note that, according to Ref.~\cite{Zeminiani:2020aho}
the $\Upsilon$ self-energy with the heavier $B^*\bar{B}^*$ (VVV coupling, V: vector meson)
gives unexpectedly large contribution, and we made our ''minimal''
predictions by avoiding this VVV coupling for the $\Upsilon$
self-energy. We follow this practice, and we do not include possible
$B^*\bar{D}^*$ ($\bar{B}^*D^*$) process in the self-energy for $B_c^*$.
This will also enable us to compare with the present
results with those of the $\Upsilon$ and $J/\Psi$ based on the same footing.
Note that, clearly no VVV coupling for $\eta_{b,c}$,
since they are pseudoscalar mesons, P.

\section{Interaction Lagrangian \label{Lint1}}

In  the relevant self-energy loop graphs for the study,
the vertex has a PPV interaction structure,
which respects the parity conservation  for the strong interaction.

To evaluate the self-energy loop graphs shown in Figs.~\ref{psmeson}
and~\ref{vmeson}, we start from a free
SU(5) symmetric effective Lagrangian in the quark flavor space to obtain the interaction Lagrangian,
and from there, we can compute the self-energies of the corresponding mesons.

The free Lagrangian of pseudoscalar (P) and vector (V) mesons based on
the flavor SU(5) symmetry is given by~\cite{Lin:2000ke,Akram:2013dhd}:
\begin{equation}
    \mathcal{L}_{0} = Tr(\partial_\mu P^{\dagger} \partial^\mu P)
    - \frac{1}{2} Tr(F^{\dagger}_{\mu\nu} F^{\mu\nu}) 
\nonumber
\end{equation}
where $F_{\mu\nu}=\partial_{\mu} V_{\nu} - \partial_{\nu} V_{\mu}$, and  
P and $V_{\mu}$ are expressed by $5 \times 5$ matrices which are obtained by using the 
generators of SU(5) (see Appendix B)  as $P^{i}_{j} = 1/\sqrt2
(\lambda_{a})_{j}^{i}\phi_{a}$ and $V_{\mu
 j}^{i} = 1/\sqrt2 (\lambda_{a})_{j}^{i}v_{\mu a}$. Here  $\phi_a$ is the Cartesian component
of pseudoscalar meson fields $(a=1,...,24)$ entering in the expression $P^{i}_{j}$,
and $v_{\mu a}$ is again
the Cartesian component of the vector meson fields entering
$V^{i}_{\mu j}$, and $\lambda_a$ are the
SU(5) version of ''Gell-Mann matrices''
$(a=1,...,24)$, and $i$ and $j$ are matrix indices (see Appendix B).

Then P and V (suppressing Lorentz indices) are expressed by the physical meson fields:
\begin{eqnarray}
\hspace{-10ex}&&\hspace{-4ex} P = \frac{1}{\sqrt{2}} \begin{pmatrix} 
\frac{\pi^{0}}{\sqrt{2}} + \frac{\eta}{\sqrt{6}} + \frac{\eta_{c}}{\sqrt{12}} 
+ \frac{\eta_{b}}{\sqrt{20}}  &  \pi^{+}  &  K^{+}  &  \overline{D}^{0}  &  B^{+}\\
\pi^{-}  &  \frac{-\pi^{0}}{\sqrt{2}} + \frac{\eta}{\sqrt{6}} + \frac{\eta_{c}}{\sqrt{12}} 
+ \frac{\eta_{b}}{\sqrt{20}}  &  K^{0}  &  D^{-}  &  B^{0}\\
K^{-}  &  \overline{K}^{0}  &  \frac{-2\eta}{\sqrt{6}} + \frac{\eta_{c}}{\sqrt{12}} 
+ \frac{\eta_{b}}{\sqrt{20}}  &  D_{s}^{-}  &  B_{s}^{0}\\
D^{0}  &  D^{+}  &  D_{s}^{+}  &  \frac{-3\eta_{c}}{\sqrt{12}} 
+ \frac{\eta_{b}}{\sqrt{20}}  &  B_{c}^{+}\\
B^{-}  &  \overline{B^{0}}  &  \overline{B_{s}^{0}}  &  B_{c}^{-}  &  \frac{-2\eta_{b}}{\sqrt{5}}\\
\end{pmatrix}, 
\hspace{8ex}\label{p1} 
\end{eqnarray}
\begin{eqnarray}
\hspace{-10ex}&&\hspace{-4ex} V = \frac{1}{\sqrt{2}} \begin{pmatrix}
\frac{\rho^{0}}{\sqrt{2}} + \frac{\omega}{\sqrt{6}} + \frac{J/\Psi}{\sqrt{12}} 
+ \frac{\Upsilon}{\sqrt{20}}  &  \rho^{+}  &  K^{*+}  &  \overline{D}^{*0}  &  B^{*+}\\
\rho^{-}  &  \frac{-\rho^{0}}{\sqrt{2}} + \frac{\omega}{\sqrt{6}} + \frac{J/\Psi}{\sqrt{12}} 
+ \frac{\Upsilon}{\sqrt{20}}  &  K^{*0}  &  D^{*-}  &  B^{*0}\\
K^{*-}  &  \overline{K}^{*0}  &  \frac{-2\omega}{\sqrt{6}} + \frac{J/\Psi}{\sqrt{12}} 
+ \frac{\Upsilon}{\sqrt{20}}  &  D_{s}^{*-}  &  B_{s}^{*0}\\
D^{*0}  &  D^{*+}  &  D_{s}^{*+}  &  \frac{-3J/\Psi}{\sqrt{12}} + \frac{\Upsilon}{\sqrt{20}}  &  
B_{c}^{*+}\\
B^{*-}  &  \overline{B^{*0}}  &  \overline{B_{s}^{*0}}  &  B_{c}^{*-}  &  
\frac{-2\Upsilon}{\sqrt{5}} \\ 
\end{pmatrix}. 
\hspace{8ex}\label{v1}
\end{eqnarray}
where we use the following conventions,
\begin{eqnarray}
K= \begin{pmatrix}
K^{+}  \\  K^{0}  
\end{pmatrix} =
\begin{pmatrix}
u  \overline{s}  \\  d  \overline{s}  
\end{pmatrix} \nonumber
,\hspace{5ex}
\overline{K} \equiv K^{\dagger} = \begin{pmatrix}
K^{-} & \overline{K}^{0} 
\end{pmatrix} =
\begin{pmatrix}
\overline{u}  s  &  \overline{d}  s  
\end{pmatrix}
\nonumber
,\end{eqnarray}

\begin{eqnarray}
\overline{D}= \begin{pmatrix}
\overline{D}^{0}  \\  D^{-}  
\end{pmatrix} =
\begin{pmatrix}
u  \overline{c}  \\  d  \overline{c}  
\end{pmatrix} \nonumber
,\hspace{5ex}
D \equiv (\overline{D})^{\dagger} = \begin{pmatrix}
D^{0} & D^{+} 
\end{pmatrix} =
\begin{pmatrix}
\overline{u}  c  &  \overline{d}  c  
\end{pmatrix}
\nonumber
,\end{eqnarray}  

\begin{eqnarray}
B= \begin{pmatrix}
B^{+}  \\  B^{0}  
\end{pmatrix} =
\begin{pmatrix}
u  \overline{b}  \\  d  \overline{b}  
\end{pmatrix} \nonumber
,\hspace{5ex}
\overline{B} \equiv B^{\dagger} = \begin{pmatrix}
B^{-} & \overline{B}^{0}
\end{pmatrix} =
\begin{pmatrix}
\overline{u}  b  &  \overline{d}  b  
\end{pmatrix}
.
\nonumber
\end{eqnarray}

Note that, in addition, similar conventions are used for
the vector mesons $K^{*}$, $D^{*}$, and $B^{*}$,
each corresponding to $K, D$ and $B$ above.

Transformation of these tensors $P^i_j$ and $V^i_j$ under SU(5) transformation
are given by,
\begin{equation}
   P^{i}_{j} \rightarrow U_{ia} P^{a}_{b} U^{\dagger b}_{j},
\end{equation}
and
\begin{equation}
    V^{i}_{j} \rightarrow U_{ia} V^{a}_{b} U^{\dagger b}_{j},
\end{equation}
where, {$P^a_b$} and $V_{\mu b}^{a}$ are the matrix representations of
the pseudo-scalar and the vector mesons
(irreducible tensors), and they are obtained using
the SU(5) generators with adjoint representation (matrix representation/notation)
by rewriting from the Cartesian components to physical components.
(See Appendix B.)

\section{Self-energy Calculation}

By the minimal substitutions~\cite{Lin:2000ke}:
\begin{equation}
    \partial_{\mu}P \rightarrow D_{\mu}P = \partial_{\mu}P - \frac{ig}{2}[V_{\mu},P], 
\end{equation}
\begin{equation}
    F_{\mu \nu} \rightarrow F_{\mu \nu} - \frac{ig}{2}[V_{\mu},V_{\nu}], 
\end{equation}
we get the interaction Lagrangian:
\begin{equation}
   \mathcal{L} = \mathcal{L}_{0} +
   igTr(\partial^{\mu} P[P,V_{\nu}] - \frac{g^{2}}{4}Tr([P,V{\mu}]^{2}) + 
igTr(\partial^{\mu}V^{\nu}[V_{\mu},V_{\nu}]) + \frac{g^{2}}{8}Tr([V_{\mu},V_{\nu}]^{2}).
\label{Lint}
\end{equation}

In Eq.~(\ref{Lint}) $\mathcal{L}_0$ represents the free Lagrangian, while the interaction terms 
describe the pseudoscalar(P)-pseudoscalar(P)-vector(V) (PPV),
PPVV, VVV and VVVV couplings. For the self-energy diagrams
shown in Figs.~\ref{psmeson} and~\ref{vmeson}, only the second term in Eq.~(\ref{Lint}),
PPV coupling, is necessary.

For example, the second term of $\mathcal{L}_{B_{c}^{*}BD}$ can be obtained
performing  $Tr{\partial_\mu}P(PV^{\mu}-V^{\mu}P)$
but, to do that, we use the P and V matrices, keeping only the terms that are involved in the
interaction:
\begin{eqnarray}
&&\hspace{-4ex} P = \frac{1}{\sqrt{2}} \begin{pmatrix} 
0 &  0 &  0  & \overline{D}^{0} & B^{+} \\
0  &  0  &  0  &  D^{-}  &  B^{0}\\
0  &  0  &  0  &  0  &  0\\
D^{0} &  D^{+} &  0  &  0  &  0\\
B^{-}  & \overline{B^{0}}  &  0 &  0 &  0 \\
\end{pmatrix}, \nonumber
	\\
\nonumber\\
\nonumber\\
&&\hspace{-4ex} V = \frac{1}{\sqrt{2}} \begin{pmatrix}
0  &  0  &  0  &  0  &  0\\
0  &  0  & 0   &  0  &  0\\
0  &  0  &  0  &  0  &  0\\
0  &  0  &  0  &  0 & B_{c}^{*+}\\
0  &  0  &  0  &  B_{c}^{*-} & 0 \\ 
\end{pmatrix}. \nonumber
\end{eqnarray}

Then, $Tr{\partial_\mu}P(PV^{\mu}-V^{\mu}P)$ gives:
\begin{equation}
    \begin{split}
        Tr{\partial_\mu}P(PV^{\mu}-V^{\mu}P) = [- (\partial_\mu \overline{D}^{o} B_c^{*+} B^{-}) -
        (\partial_\mu B^{+} B_c^{*-} D^{0}) ] \\
        + [ -(\partial_\mu 
        {D}^{-} B_c^{*+} \overline{B}^{0}) - (\partial_\mu B^{0} B_c{*-} D^{+}) ] \\
        + [ (\partial_\mu D^{0} B^{+} B_c^{*-}) + (\partial_\mu D{+} B^{0} B_c^{-})] \\
        + [(\partial_{\mu} B^{-} \overline{D}^{0} B_c^{*+}) + (\partial_\mu \overline{B}^{0} D^{-}
B_c^{*+}) ].
    \end{split}
\label{trace}
 \end{equation}

Performing the trace in the above,
we get the interaction Lagrangian for $B_c$:
\begin{equation}
     \mathcal{L}_{B^{*}_{c}BD} = -ig_{B^{*}_{c}BD}[B(\partial_{\mu} \overline{D}) -
(\partial_{\mu}B) \overline{D}]B^{* {\mu}}_{c} +  h.c.
\nonumber
\end{equation}

Finally, for the set of interaction terms required to study the interactions
of $B_{c}$, $B_s$, and $D_s$ and $B^{*}_{c}$, $B^{*}_{s}$ and $D^{*}_{s}$ we get:
\bea
    &\mathcal{L}_{B_{c}B^{*}D} = ig_{B_{c}B^{*}D}
    [(\partial_{\mu}B^{-}_{c}){D}
    - B^{-}_{c}(\partial_{\mu}{D})] B^{* \mu} + h.c.
\\
    &\mathcal{L}_{B_{c}BD^{*}} = ig_{B_{c}BD^{*}}
    [(\partial_{\mu}B^{+}_{c})\overline{B}
    - B_c^{+}(\partial_{\mu}\overline{B})] \overline{D^{*}}^\mu + h.c.
\\
     &\mathcal{L}_{B_sB^{*}K} = ig_{B_sB^{*}K}[(\partial_{\mu}\overline{B^0_s})\overline{K}
     - \overline{B^0_s} (\partial_{\mu}\overline{K})]{B}^{* \mu} + h.c.
\\
     &\mathcal{L}_{B_sBK^*} = ig_{B_sBK^*}[(\partial_{\mu}{B^0_s})\overline{B}
     - {B^0_s}(\partial_{\mu}\overline{B})]{K}^{* \mu} + h.c.
\\
    &\mathcal{L}_{D_sD^{*}K} = ig_{D_sD^{*}K}[(\partial_{\mu}D_{s}^+)\overline{K}
    - D_s^+ (\partial_\mu \overline{K})]\overline{D^*}^\mu + h.c.
\\
    &\mathcal{L}_{D_sDK^*} = ig_{D_sDK^*}[(\partial_{\mu}D^-_{s})D
    - D_{s}^- (\partial_{\mu}D)  ]{K}^{* {\mu}} + h.c.
\\
     &\mathcal{L}_{B^{*}_{c}BD} = -ig_{B^{*}_{c}BD}
     B^{*+{\mu}}_{c} [\overline{B}(\partial_{\mu} \overline{D}) -
(\partial_{\mu} \overline{B}) \overline{D}] + h.c.
\\
     &\mathcal{L}_{B^{*}_{s}BK} = -ig_{B^{*}_{s}BK}
     {B_s^*}^{0 \mu} [\overline{B} (\partial_\mu K)
     - (\partial_\mu \overline{B}) K]  + h.c.
\\
     &\mathcal{L}_{D^{*}_{s}DK} = -ig_{D^{*}_{s}DK}
     {D_s^{*-}}^\mu [D(\partial_{\mu}K)
     - (\partial_{\mu}D) K]  + h.c.
\eea

With these interaction Lagrangians,
we can calculate self-energy graphs, that results in
the Lorentz scalar effective meson mass, or potential $V_s$,
for the corresponding mesons.
Note that, for the quarks $s, c$ and $b$, the present approach
assumes that there arises no ''Lorentz vector'', in-medium potentials based on the QMC model.
(See Ref.~\cite{Guichon:1987jp,Stone:2016qmi,Guichon:2018uew}.)


Based on the vector meson dominance (VMD) hypothesis and
experimental data of $\Gamma(\Upsilon\rightarrow e^{+}e^{-})$ 
we have calculated the coupling constant  $g_{\Upsilon BB}$~\cite{Lin:2000ke}
(see Appendix C).

\begin{equation}
g_{\Upsilon BB} = g_{\Upsilon B^{*}B} = \frac{5g}{4\sqrt{10}} \approx 13.2
\label{eq36}
\end{equation}

Using Eq.~(\ref{eq36}) ~\cite{Zeminiani:2021xvw,Zeminiani:2020aho}
we obtain $g \approx 33.45$,
and  the  relevant coupling constants by:
\bea
    &&g_{B_cBD^{*}} = g_{B_cB^{*}D} = g_{B_c^*BD} =
    \frac{g}{2\sqrt{2}},
\\
    &&g_{B_cBD^{*}} = g_{B_cB^{*}D} =
    g_{B_cB_sD_s^{*}} =
    g_{B_cB_s^{*}D_s} =
    \frac{g}{2\sqrt{2}},
\\
    &&g_{KD^{*}D_s} = g_{KDD_s^{*}} =
    g_{KB^{*}D_s} =
    g_{KBB_s^{*}} =
    \frac{g}{2\sqrt{2}}.
\eea
Thus, we get $g_{B_cBD^{*}} = g_{B_cB^{*}D} \approx$ 11.9

We must, however, draw attention to the coupling constant used in the present work.
To obtain the above value, we use the experimental data from the $\Upsilon$ decay, although we could instead perform a similar calculation using the experimental data from the $J/\Psi$ decay, which would lead us to obtain a different value for $g_{B_cBD^{*}} (g_{B_cB^{*}D})$.
It, of course, would imply in different numerical values for all results, but all the relative relations would remain the same.


In the following we simplify
the effective Lagrangians for $B_c$ and $B_c^*$.
In order to pay attention and emphasize some points,
there are terms with underlines in the Lagrangians throughout hereafter.
\begin{equation}
\mathcal{L}_{B_{c}B^{*}D} =
ig_{B_{c}B^{*}D} \overline{B^*}_{\mu}[\underline{B^{+}_{c}(\del^{\mu}\Dbar)} -
(\del^{\mu}B^{+}_{c})\Dbar]
-ig_{B_{c}B^{*}D}[\underline{B^{-}_{c}(\del^{\mu}D)} -  (\del^{\mu}B^{-}_{c})D]B^{*}_{\mu}
\end{equation}
\begin{equation}
     \mathcal{L}_{DBB^{*}_{c}} = 
ig_{DBB^{*}_{c}} B^{*+}_{c~\mu}[(\partial^{\mu} \Bbar)\Dbar -
\underline{\Bbar(\partial^{\mu}\Dbar)}]
-ig_{DBB^{*}_{c}}B^{*-}_{c~\mu} [\Dbar (\partial^{\mu} \overline{B)} -
\underline{(\partial^{\mu}\Dbar) \overline{B}}]
\end{equation}

Integrating by parts the underlined terms and dropping the surface terms ($= 0$),
and using Lorenz condition $\partial^\mu B^{*}_{\mu}$
and $\partial^\mu \overline{B^*}_{\mu} = 0$, we get the simplified expressions
for the $B_c$ interaction Lagrangian,
\begin{equation}
\mathcal{L}_{B_{c}B^{*}D} =  2 ig_{B_{c}B^{*}D}[-(\del^{\mu}B^{+}_{c})\Bbar^{*}_{\mu}\Dbar +
(\del^{\mu}B^{-}_{c})DB^{*}_{\mu}],
\end{equation}
and for $B_c^*$,
\begin{equation}
\mathcal{L}_{DBB^{*}_{c}} =  2 ig_{DBB^{*}_{c}}[ B^{*+}_{c\mu}(\del^{\mu}\Bbar)\Dbar -
B^{*-}_{c\mu} D(\del^{\mu}B)].
\end{equation}					

The explicit use of meson doublet fields are:
\begin{center}
$\Bbar^{*}_{\mu}\Dbar = \begin{pmatrix}
B^{*-}_{\mu} & \overline{B}^{*0}_{\mu}
\end{pmatrix} \begin{pmatrix}
\overline{D}^{0} \\ D^{-} 
\end{pmatrix} = B^{*-}_{\mu}\Dbar^{0} + \Bbar^{*0}D^{-}$,
\end{center}
\begin{center}
$DB = \begin{pmatrix}
{D}^{0} & D^{+} 
\end{pmatrix} \begin{pmatrix}
B^{+} \\ B^{0}
\end{pmatrix} = D^{0}B^{+} + D^{+}B^{0}$,
\end{center}
and
\begin{center}
$\Bbar\Dbar = \begin{pmatrix}
B^{-} & \overline{B}^{0}
\end{pmatrix} \begin{pmatrix}
\overline{D}^{0} \\ D^{-} 
\end{pmatrix} = B^{-}\Dbar^{0} + \Bbar^{0}D^{-}$,
\end{center}
\begin{center}
$DB = \begin{pmatrix}
{D}^{0} & D^{+} 
\end{pmatrix} \begin{pmatrix}
B^{+} \\ B^{0}
\end{pmatrix} = D^{0}B^{+} + D^{+}B^{0}$.
\end{center}

From the general expression for scattering matrix, $S$:
\begin{equation}
\begin{aligned}
S &= 1 - i\mathcal{T},\\
&= 1 - (2\pi)^{4} \delta^{(4)}(p_f-p_i)i\mathcal{M},\\
&= 1 + (2\pi)^{4} \delta^{(4)}(p_f-p_i)(-i\mathcal{M}),
\end{aligned}
\end{equation}
where $\mathcal{T}$ and $\mathcal{M}$ are respectively the transition matrix and invariant
amplitude, and we can define, $-i\mathcal{M} = -i\Sigma$ and, therefore
$\mathcal{M} = \Sigma$ with $\Sigma$ to be associated with the self-energy.

We calculate the lowest order non-trivial $S$ matrix (second order for the connected
processes) for $B_c$ and $B_c^*$ below.\\
For $B_{c}$,
\begin{equation}
S_{2} = <B_{c}^{+}(p)| T \frac{i^{2}}{2!} \int d^{4}xd^{4}y \mathcal{L}_{B_{c}B^{*}D}^{(x)}
\mathcal{L}_{B_{c}B^{*}D}^{(y)} |B_{c}^{+}(p)>,
\end{equation}
and for $B^{*}_{c}$,
\begin{equation}
S_{2} = <B_{c\mu}^{*+}(p)| T \frac{i^{2}}{2!}\int d^{4}xd^{4}y
\mathcal{L}_{DBB^{*}_{c}}^{(x)}\mathcal{L}_{DBB^{*}_{c}}^{(y)}|B_{c\nu}^{*+}(p)>.
\end{equation}

By exchanging the integral variables  $x \leftrightarrow y$, and summing the terms
which give the same contributions, we get the following expressions.\\
\noindent
For $B_{c}$,
\begin{equation}
S_{2} = -4g^{2}_{B_{c}B^{*}D}  \int d^{4}xd^{4}y <0| 
  (e^{ip \cdot x}) T  \underline{[\Bbar^{*}_{\mu}\Dbar]_{(x)}
[DB^{*}_{\nu}]_{(y)}}(e^{-ip \cdot x})
|0> p^{\mu}p^{\nu},
\label{eq12}
\end{equation}
and for $B^{*}_{c}$,
\begin{equation}
S_{2} = <0| 
T \int d^{4}xd^{4}y (\epsilon^{*}_{\mu}(\lambda)e^{ip \cdot x}) \cdot 4g^{2}_{DBB_{c}^{*}}
\underline{[-D(\del^{\mu}_{x}B)]_{(x)}
[(\del^{\nu}_{y}\Bbar)\Dbar]_{(y)}}(\epsilon^{*}_{\nu}(\lambda)e^{-ip \cdot x}) |0>.
\label{eq13}
\end{equation}

Expanding the underlined terms in Eqs.~(\ref{eq12}) and~(\ref{eq13}), Feynman propagators
show up as Eqs.~(\ref{eq14}),~(\ref{eq15}),~(\ref{eq16}), and~(\ref{eq17}) below.\\
\noindent
For $B_{c}$,
\begin{equation}
i\Delta_{D}(x-y) = i \int \frac {d^{4}\kappa_{1}}{(2\pi)^{4}}
\frac{e^{-i\kappa_{1}\cdot(x-y)}}{\kappa^{2}-m^{2}_{D}+i\epsilon},
\label{eq14}
\end{equation}
and
\begin{equation}
i\Delta_{B^{*}_{\mu\nu}} (x-y) = i \int \frac {d^{4}\kappa_{2}}{(2\pi)^{4}}\frac
{(-g_{\mu\nu}+\kappa_{2\mu} \kappa_{2\nu}/m^{2}_{B^{*}}) e^{-i\kappa_{2}\cdot
(x-y)}}{\kappa^{2}_{2}-m^{2}_{B^{*}}+i\epsilon}.
\label{eq15}
\end{equation}
In addition, for $B^{*}_{c}$, also appear,
\begin{equation}
i\Delta_{B}(x-y) = i\int \frac{d^{4}\kappa_{2}}{(2\pi)^{4}}
\frac{e^{-i\kappa_{2}\cdot(x-y)}}{\kappa_{2}^{2}-m^{2}_{B}+i\epsilon},
\label{eq16}
\end{equation}
and
\begin{equation}
i\Delta_{D}(x-y) = i\int \frac{d^{4}\kappa_{1}}{(2\pi)^{4}}
\frac{e^{-i\kappa_{1}\cdot(x-y)}}{\kappa_{1}^{2}-m^{2}_{D}+i\epsilon}.
\label{eq17}
\end{equation}

Inserting the above expressions into expressions for $S_{2}$,
Eqs.~(\ref{eq12}), and~(\ref{eq13}), we obtain:\\
\noindent
For $B_{c}$,
\bg
S_{2}&=& \int d^{4}xd^{4}y (-4g^{2}_{B_{c}B^{*}D}) e^{-ip \cdot y}e^{ip \cdot x} \cdot 2
(p^{\mu}p^{\nu})
\nonumber\\
& & \times~ i \int \frac {d^{4}\kappa_{1}}{(2\pi)^{4}}
\frac{e^{-i\kappa_{1}\cdot(x-y)}}{\kappa^{2}_{1}-m^{2}_{D}+i\epsilon} i \int \frac
{d^{4}\kappa_{2}}{(2\pi)^{4}}\frac {(-g_{\mu\nu}+\kappa_{2\mu} \kappa_{2\nu})/m^{2}_{B^{*}})
e^{-i\kappa_{2}\cdot (x-y)}}{\kappa^{2}_{2}-m^{2}_{B^{*}}+i\epsilon},
\en
and for $B^{*}_{c}$,
\bg
S_{2} &=& \int d^{4}xd^{4}y \epsilon^{*}_{\mu}(\lambda)\epsilon_{\nu}(\lambda)e^{ip \cdot x}e^{-ip
\cdot y}(4g^{2}_{DBB^{*}_{c}})
\nonumber\\
& &\times ~2~ \int \frac{d^{4} \kappa_{1}}{(2\pi)^{4}}
\frac{e^{-i\kappa_{1}\cdot(x-y)}}{\kappa_{1}^{2}-m^{2}_{D}+i\epsilon} \int
\frac{d^{4}\kappa_{2}}{(2\pi)^{4}}
\frac{\del^{x}_{\mu}\del^{y}_{\nu}e^{-i\kappa_{2}\cdot(x-y)}}{\kappa_{2}^{2}-m^{2}_{B}+i\epsilon}.
\en

With $p=\kappa_{1}+\kappa_{2}$ , $\kappa_{2} \equiv \kappa$ $\ra \kappa_{1} = p-\kappa$, we obtain:\\
\noindent
For $B_c$,
\begin{equation}
S_{2}= 8g^{2}_{B_{c}B^{*}D} (p^{\mu}p^{\nu}) \int \frac{d^{4}\kappa}{(2\pi)^{4}} \frac{(-g_{\mu\nu}
+\kappa_{\mu}\kappa_{\nu}/m^{2}_{B^{*}})}{\kappa^{2}-m^{2}_{B^{*}}+i\epsilon}
\frac{1}{(p-k)^{2}-m^{2}_{D}+i\epsilon} (2\pi)^{4}\delta^{(4)}(p-p),
\end{equation}
and for $B_c^*$,
\begin{equation}
\begin{aligned}
S_{2} 
= 8g^{2}_{B^{*}_{c}BD} \int
\frac{d^{4}\kappa}{(2\pi)^{4}}\kappa^{\mu}\kappa^{\nu}\epsilon^{*}_{\mu}(\lambda)\epsilon_{\nu}
(\lambda)\frac{1}{\kappa^{2}-m^{2}_{B}+i\epsilon}\frac{1}{(p-\kappa)^{2}-m^{2}_{D}+i\epsilon}(2\pi)
^{ 4}\delta^{(4)}(p-p)
\end{aligned}
\end {equation}

Since, $-i\mathcal{M} = -i\Sigma$, then we obtain: \\
For $B_{c}$,
\begin{equation}
\Sigma^{B^{*}D}_{B_{c}}(p) = -\frac{1}{i}8g^{2}_{B_{c}B^{*}D} (p^{\mu}p^{\nu}) \int
\frac{d^{4}\kappa}{(2\pi)^{4}}\frac
{(-g_{\mu\nu}+\kappa_{\mu}\kappa_{\nu}/m^{2}_{B^{*}})}{\kappa^{2}-m^{2}_{B^{*}}+i\epsilon}
\frac{1}{(p-\kappa)^{2}-m^{2}_{D}+i\epsilon}(2\pi)^{ 4}\delta^{(4)}(p-p),
\label{BcSEeq} 
\end{equation}
and for $B^{*}_{c}$,
\begin{equation}
\Sigma^{BD}_{B^{*}_{c}} (p) = -\frac{1}{i}8g^{2}_{DBB^{*}_{c}} \int
\frac{d^{4}\kappa}{(2\pi)^{4}}\kappa^{\mu}\kappa^{\nu}\epsilon^{*}_{\mu}(\lambda)\epsilon_{\nu}
(\lambda)\frac{1}{\kappa^{2}-m^{2}_{B}+i\epsilon}\frac{1}{(p-\kappa)^{2}-m^{2}_{D}+i\epsilon}
(2\pi)^{ 4}\delta^{(4)}(p-p).
\label{BcsSEeq}   
\end{equation}

Note that, although we estimate the mass shift of $B^{*}_{c}$ meson,
the existence is suggested by theoretically and
the existence has not yet been confirmed by experiment.
We emphasize that, our study brings the first predictions for the mass shift
of both the $B_c$ and $B_c^*$ mesons in a nuclear medium (many nucleon system).

Similar calculations can also be done for $B_{s}$, $D_{s}$ pseudo-scalar
mesons, as well as for $B_{s}^{*}$, and $D_{s}^{*}$ vector mesons.
Of course, all the estimates for the mass shift of these mesons are
our predictions, and are made for the first time, which means, no one has ever
calculated/estimated.




We first calculate the free space physical meson mass,
\begin{equation}
m^{2} = (m^{0})^{2} - |\Sigma(p^{2}=m^{2})|,
\label{m0}
\end{equation}
where $m^{0}$ is the bare mass, $p$ is the meson four-momentum, $\Sigma$ the total self-energy in free space which is the sum of each loop contribution in free space.
(See Figs.~\ref{psmeson} and~\ref{vmeson}.)
To make clear the sign convention, we introduced the modulus as
$|\Sigma|$, based on the fact that ''the second-order perturbation gives always
the negative contribution for the energy (mass) with the virtual state excitations''.
According to Eq.~(\ref{m0}), we fix the $m^0$ value by the free space physical meson mass $m$.
To calculate the in-medium meson mass $m^*$, we use this fixed $m^0$ value,
together with the in-medium self-energy $\Sigma^*$ corresponding to Eq.~(\ref{m0}).

\noindent

In  symmetric nuclear matter rest frame,
we consider the $B_{c}$ meson at rest,
$p^{\mu}=(m^{*}_{B_{c}}, \vec 0)$.
The numerator $(-g_{\mu\nu}+\kappa_{\mu}\kappa_{\nu})/m^{*2}_{B^*}$ of $B^{*}$
meson propagator in Eq.~(\ref{BcSEeq}) can then be evaluated in this rest frame:
\begin{equation}
\begin{aligned}
p^{\mu}p^{\nu}(-g_{\mu\nu}+\kappa_{\mu}\kappa_{\nu})/m^{*2}_{B^*}
&= -p^{2}+(p\cdot \kappa ) (p\cdot \kappa) / m^{*2}_{B^*},\\
&= -m^{*2}_{B_{c}}+(m^{*}_{B_{c}}\kappa^{*}_{0})(m^{*}_{B_{c}}\kappa^{*}_{0}) / m^{*2}_{B^*},\\
&= m^{*2}_{B_{c}}(-1+ \kappa^{*2}_{0}/m^{*2}_{B^{*}}),\\
&= (\frac{m^{*2}_{B_{c}}}{m^{*2}_{B^{*}}})(-m^{*2}_{B^{*}}+k^{*2}_{0}),\\
&=(\frac{m^{*2}_{B_{c}}}{m^{*2}_{B^{*}}})((\kappa_{0}+{V})^{2}-m^{*2}_{B^{*}}),\\
&=(\frac{m^{*2}_{B_{c}}}{m^{*2}_{B^{*}}})(\kappa_{0}+V+m^{*}_{B^{*}})
(\kappa_{0}+V-m^{*}_{B^{*}}).
\end{aligned}
\label{Bsprop}
\end{equation}


In the above, $V$ stands for the time component of the vector-mean filed potential
which comes from the QMC model approach, and adds to the time component
of the in-medium energy as $\kappa_0^* = \kappa_0 + V$.
See Refs.~\cite{Saito:2005rv,Tsushima:1997df,Tsushima:2019wmq}
and Appendix B for details.
As for $B_{c}^{*}$ meson, $B_{c}^{*}$ since is a vector meson, its spin average is $1/(2J+1) = 1/3$,
and in the rest frame of $B_{c}^{*}$ (in free space),
since $p^{\mu}=(m_{B_{c}^{*}}, \vec 0)$, we get,
\bea
\frac{1}{3}\sum\limits_{spins}\kappa^{\mu}\kappa^{\nu}\epsilon^{*}_{\mu}(\lambda)\epsilon_{\nu}
(\lambda)
&=&
\frac{1}{3}(-g_{\mu\nu}+\frac{p_{\mu}p_{\nu}}{m^{2}_{B_{c}^{*}}})\kappa^{\mu}\kappa^{\nu},
\nonumber\\
&=& \frac{1}{3}(-\kappa^{2}+ \frac{1}{m^{2}_{B^{*}_{c}}}(p_{0}k^{0})(p_{0}k^{0})),
\nonumber\\
&=& \frac{1}{3}(-\kappa^{2}+ \frac{1}{m^{2}_{B^{*}_{c}}}m^{2}_{B^{*}_{c}}(k^{0})^{2}),
\nonumber\\
&=& \frac{1}{3}(-(\kappa^{0})^{2}+\vec \kappa^{2} + (\kappa^{0})^{2}),
\nonumber\\
&=& \frac{1}{3}|\vec \kappa^{2}|.
\eea
where in the first term, it corresponds to $\frac{1}{2J+1}\sum \limits_{spins}$.
Furthermore,
\begin{equation}
\begin{aligned}
\int \frac{d^{4}\kappa}{(2\pi)^{3}}&= \int \frac{d^{3}\kappa}{(2\pi)^{3}} \int
\frac{d\kappa^{0}}{2\pi},\\
&=\int \frac{d|\vec \kappa|4\pi|\vec \kappa|^{2}}{(2\pi)^{3}} \int \frac{dk^{0}}{2\pi},\\
&= \frac{1}{2\pi^{2}} \int d|\vec \kappa| |\vec \kappa|^{2} \int \frac{d\kappa^{0}}{2\pi}.
\end{aligned}
\end{equation}
Thus, we get the expressions for $B_{c}$ and for $B^{*}_{c}$:
\bea
\Sigma^{B^{*}B}_{B_{c}}(m_{B_{c}}) &=& \frac{-4g^{2}_{B_{c}B^{*}D}}{\pi^{2}}
\int d |\vec \kappa||\vec \kappa|^{2}
\nn\\
& & \hspace{-10ex} \times \int \frac{d{\kappa_{0}}}{2\pi i}
(\frac{m^{*2}_{B_{c}}}{m^{2}_{B^{*}}})
\frac{(\kappa_{0}+V+m_{B^{*}})(\kappa_{0}+V-m_{B^{*}})}{\kappa^{2}-m^{2}_{B^{*}}+i\epsilon}
\frac{1}{(p-\kappa)^2-m^{2}_{D}+i\epsilon},
\eea
,
\begin{equation}
\Sigma^{BD}_{B^{*}_{c}}(m_{B^{*}_{c}}) = \frac{-4g^{2}_{DBB^{*}_{c}}}{3\pi^{2}} \int d|\vec \kappa| |\vec \kappa|^{4} 
\int \frac{d\kappa^{0}}{2\pi i} 
\frac{1}{\kappa^{2}-m^{2}_{B}+i\epsilon}\frac{1}{(p-\kappa)^{2}-m^{2}_{D}+i\epsilon}.
\end{equation}
Therefore the in-medium self-energies are given by (the relevant meson masses are replaced by those of in medium):
\\
For $B_{c}$,
\bg
\Sigma^{B^{*}B}_{B_{c}}(m^{*}_{B_{c}})
&=& \frac{-4g^2_{B_cB^*D}}{\pi^2} \int d |\vec{\kappa}| |\vec{\kappa}|^{2}
\nonumber\\
& & \hspace{-10ex} \times \int \frac{d \kappa_0}{2\pi i} (\frac{m^{*2}_{B_c}}{m^{*2}_{B^*}})
\frac{(\kappa_0+V+m^*_{B^*})(\kappa_0+V-m^*_{B^*})}{\kappa^2-m^{*2}_{B^*}+i\epsilon}
 \frac{1}{(p-\kappa)^2-m^{*2}_{D}+i\epsilon},
\en
and for $B_c^*$,
\begin{equation}
\Sigma^{BD}_{B^{*}_{c}}(m^{*}_{B^{*}_{c}}) = \frac{-4g^{2}_{ B^{*}_{c}BD}}{3\pi^{2}} \int
d|\vec \kappa| |\vec \kappa|^{4}
\int \frac{d\kappa^{0}}{2\pi i} 
\frac{1}{\kappa^{2}-m^{*2}_{B}+i\epsilon}\frac{1}{(p-\kappa)^{2}-m^{*2}_{D}+i\epsilon}.
\end{equation}


In  order to evaluate the poles in the integral, we define:
\\
\noindent
For $B_{c}$,
\bge
I_{B_c}^{B^{*}D} = (\frac{m^{*2}_{B_{c}}}{m^{*2}_{B^{*}}})  \int \frac{d{\kappa_{0}}}{2\pi i}
\frac{(\kappa_{0}+V+m^*_{B^{*}})(\kappa_{0}+V-m^*_{B^{*}})}{\kappa^{2}-m^{* 2}_{B^{*}}+i\epsilon}
\frac{1}{(p-\kappa)^2-m^{* 2}_{D}+i\epsilon},
\label{BcInt}
\ene
and for $B^{*}_{c}$,
\bge
I^{BD}_{B^{*}_{c}} = \int \frac{1}{\kappa^{2}-m^{*2}_{B}+i\epsilon}
\frac{1}{(p-k)^{2}-m^{*2}_{D}+i\epsilon}.
\label{BcsInt}
\ene










Inserting  $\kappa_{0}= -\omega^{*}_{B}$ and $\kappa_{0}= m^{*}_{B_{c}}-\omega^{*}_{D}$,
and performing the Cauchy integral in the complex plane for $\kappa_0$,
$I_{B_c}^{B^{*}D}$ and $I^{BD}_{B^{*}_{c}}$ become:
\\
\noindent
For $B_{c}$,
\bea
I_{B_c}^{B^{*}D}(|\vec{\kappa}|) &\equiv& (\frac{m^{2}_{B_{c}}}{m^{2}_{B^{*}}})  \cdot
[\frac{(-\omega^{*}_{B^{*}}+m^{*}_{B^{*}})(-\omega^{*}_{B^{*}}-m^{*}_{B^{*}})}{(-2\omega^{*}_{B^{*}}
)(-\omega^{*}_{B^{*}}-m^{*}_{B_{c}}+\omega^{*}_{D})(-\omega^{*}_{B^{*}}-m^{*}_{B_{c}}-\omega^{*}_{D}
)}
\nonumber\\
& & \hspace{10ex} +\frac{(m^{*}_{B_{c}}-\omega^{*}_{D}+m^{*}_{B^{*}})(m^{*}_{B_{c}}
-\omega^{*}_{D}-m^{*}_{B^{*}})}{(
m^{*}_{B_{c}}-\omega^{*}_{D}+\omega^{*}_{B^{*}})(m^{*}_{B_{c}}-\omega^{*}_{D}-\omega^{*}_{B^{*}})(-2
\omega^{*}_{D})}],
\eea
and for $B^{*}_{c}$,
\bea
I^{BD}_{B^{*}_{c}} (|\vec\kappa|) &=&
\frac{1}{ (-2\omega^{*}_{B}) (-\omega^{*}_{B}-m^{*}_{B^{*}_{c}}+\omega^{*}_{D})
(-\omega^{*}_{B}-m^{*}_{B^{*}_{c}}-\omega^{*}_{D}) }
\nn\\
& &\hspace{5ex} +  \frac{1}{(m^{*}_{B^{*}_{c}}-\omega^{*}_{D}+\omega^{*}_{B})
(m^{*}_{B^{*}_{c}}-\omega^{*}_{D}-\omega^{*}_{B}) (-2\omega^{*}_{D})}.
\eea


Finally. the self-energies of $B_c$ and $B_c^*$ can be expressed as:\\
\noindent
For $B_{c}$,
\begin{equation}
\Sigma^{B^{*}D}_{B_{c}}
(m^{* 2}_{B_{c}})
= \frac{-4g^{2}_{B_{c}B^{*}D}}{\pi^{2}} \int d|\vec
\kappa|
|\vec \kappa|^{2} I_{B^{*}D}(|\vec \kappa|),
\end{equation}
and for $B^{*}_{c}$,
\begin{equation}
\Sigma^{BD}_{B^{*}_{c}}
(m^{* 2}_{B_c^*})
= \frac{-4g^{2}_{B^{*}_{c}BD}}{3\pi^{2}} \int
d|\vec\kappa|
|\vec\kappa| ^{4} I^{BD}_{B^{*}_{c}}(|\vec\kappa|).
\end{equation}

It is straightforward to repeat the processes shown above to find
the self-energies of pseudoscalar
mesons, $B_{c}$, $B_{s}$, and $D_{s}$,  as well as of
vector mesons $B^{*}_{c}$, $B^{*}_{s}$, and $D^{*}_{s}$.

Note that the total the self-energy $\Sigma_{B_c}$ is calculated summing 
both loop contributions in free space
and ignoring the possible width (imaginary part) in self-energy.
 
Similar to Eq.~(\ref{m0}) the in-medium mass is calculated
using the same bare mass value $m^{0}_{B_{c}}$ determined in free space with the in medium masses, for example,

\begin{equation}
m^{2}_{B_c} = [m^{0}_{B_c} (B^{*}D + BD^{*})]^{2} -
|\Sigma_{B_c} (B^{*}D) + \Sigma_{B_c} (BD^{*})| (\kappa^{2}=m^{2}_{B_c}),
\label{mbc}
\end{equation}
\begin{equation}
m^{*2}_{B_c} = [m^{0}_{B_c} (B^{*}D + BD^{*})]^{2} -
|\Sigma_{B_c} (B^{*}D) + \Sigma_{B_c} (BD^{*})| (\kappa^{*2}=m^{*2}_{B_c}).
\label{msbc}
\end{equation}

When we consider each loop contribution $\Sigma_{B_c}  = \Sigma(B^{*}D)$ + $\Sigma(BD^{*})$, 
$m^{0}$ is a different value from that it is calculated in a situation considering only one loop.

In other words $m^{0}(B^{*}D + BD^{*}) \ne m^{0}(B^{*}D)$, $ m^{0}(B^{*}D + BD^{*}) \ne m^{0}(BD^{*})$
and nor $m^{0}(B^{*}D + BD^{*}) \ne m^{0}(B^{*}D) + m^{0}(BD^{*})$.
Thus, one should not compare
directly the each loop contribution when the in-medium mass shift will be presented later.

From each self-energy contribution, one can
only know which self-energy contribution is more dominant for the
in-medium mass shift. 

Each contribution reflects the mass reduction ''effectiveness'' for the total in-medium mass shift.

\chapter{Results}

\section{$B_{c}$ and $B^{*}_{c}$ meson results}

Before presenting the numerical results (our predictions),
it should be commented that, a phenomenological vertex form factor is used
for each interaction vertex to tame the divergent loop integral
appearing in the self-energy calculation.
More about the regularization and the methods can be found in
Ref.~\cite{Huang:2013zaa,tHooft:1972tcz,Bollini:1972ui,Cicuta:1972jf}.

For the $B_{c}$ pseudoscalar meson we use the following form factors

\bg
u_{B^{*}DB^{*}}(\vec{q}^{\,2}) = \left(\frac{\Lambda^{2}_{B^{*}D} + m^{2}_{B_{c}}}{\Lambda^{2}_{B^{*}D}
+
4\omega^{2}_{B^{*}} (\vec{q}^{\,2})}\right)^{2},
\nonumber
& &u_{B^{*}DD}(\vec{q}^{\,2}) = \left(\frac{\Lambda^{2}_{B^{*}D} + m^{2}_{B_{c}}}{\Lambda^{2}_{B^{*}D} +
4\omega^{2}_{D} (\vec{q}^{\,2})}\right)^{2},
\nonumber\\
\vspace{5ex}
u_{BD^{*}D^{*}}(\vec{q}^{\,2}) = \left(\frac{\Lambda^{2}_{BD^{*}} + m^{2}_{B_{c}}}{\Lambda^{2}_{BD^{*}} +
4\omega^{2}_{D^{*}} (\vec{q}^{\,2})}\right)^{2},
\nonumber
& &u_{BD^{*}B}(\vec{q}^{\,2}) = \left(\frac{\Lambda^{2}_{BD^{*}} + m^{2}_{B_{c}}}{\Lambda^{2}_{BD^{*}} +
4\omega^{2}_{B} (\vec{q}^{\,2})}\right)^{2}.
\nonumber
\en

And for the $B^{*}_{c}$ vector meson we use:
\bg
u_{BDB}(\vec{q}^{\,2}) = \left(\frac{\Lambda^{2}_{BD} + m^{2}_{B_{c}^{*}}}{\Lambda^{2}_{BD} +
4\omega^{2}_{B} (\vec{q}^{\,2})}\right)^{2},
\nonumber\\
\vspace{5ex}
u_{BDD}(\vec{q}^{2}) = \left(\frac{\Lambda^{2}_{BD} + m^{2}_{B^{*}_{c}}}{\Lambda^{2}_{BD} +
4\omega^{2}_{D} (\vec{q}^{\,2})}\right)^{2}.
\nonumber
\en

In general,
form factors are used to phenomenologically describe the
finite size effect of particles
and their interaction ranges or effectiveness.
Thus, the form  factors are necessary to include the effects of
the finite sizes of the mesons and the
overlapping regions associated with the interaction vertices.
For this work we use dipole form factor which was well successful in
previous works~\cite{Krein:2013rha,Tsushima:2011kh,Cobos-Martinez:2017vtr,Cobos-Martinez:2017woo,
Cobos-Martinez:2017onm,Cobos-Martinez:2017fch}.

The form factors contain the cutoff parameters,
denoted by $\Lambda_{B^*D}, \Lambda_{BD^*}$ and $\Lambda_{BD}$
which play non-negligible role to regularize the divergent integral
and to get the final results.
Note that, we will use,
$\Lambda_{B^*D} = \Lambda_{BD^*} = \Lambda_{BD} \equiv
\Lambda$ hereafter.
The value can introduce systematic uncertainties,
and different physical processes might require different
cutoff mass values, and possibly different form factors.
For all the mesons studied in this thesis
we use five different cutoff values varying
$\Lambda$ = 2000, 3000, 4000, 5000, 6000 MeV.
More details on the form factors can be seen in
Refs.~\cite{Zeminiani:2021xvw,Zeminiani:2021vaq,Zeminiani:2020aho}.

\begin{figure}[htb!]
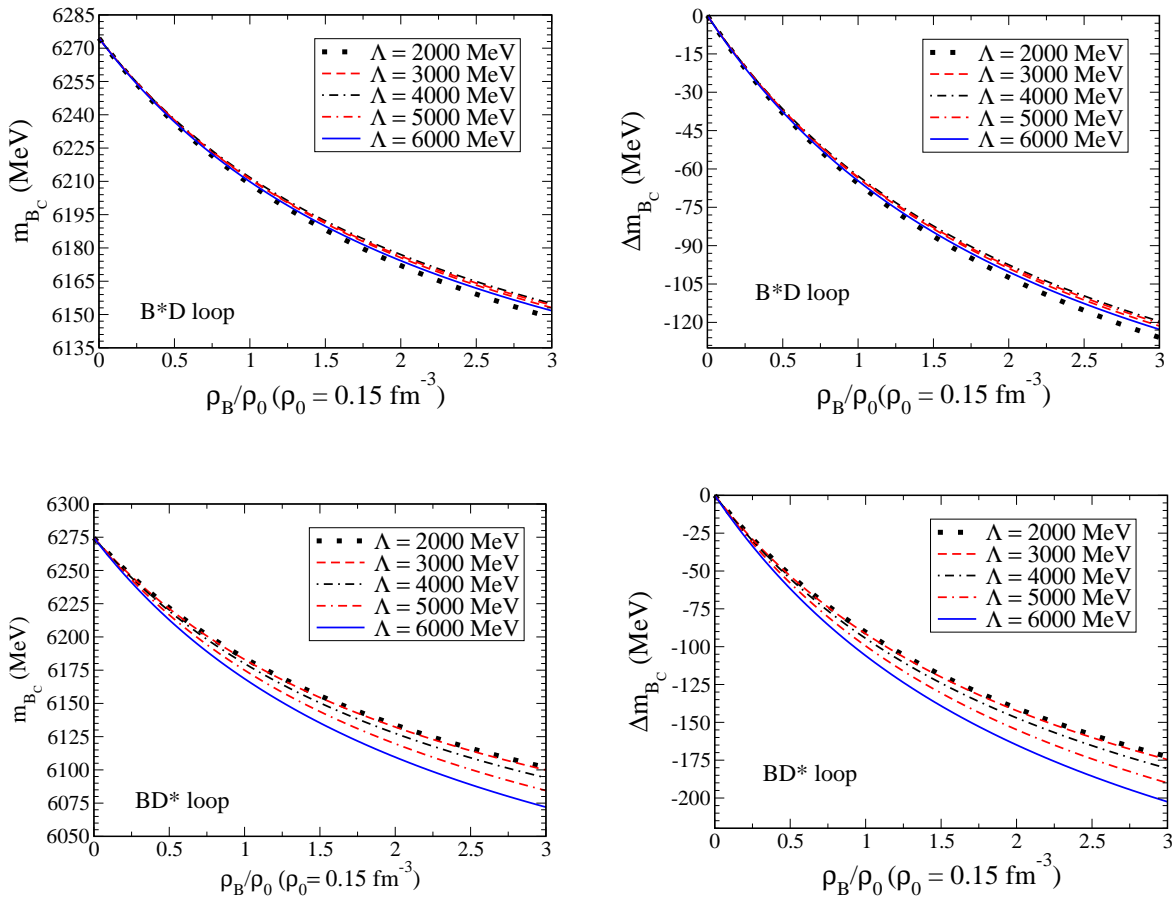

\vspace{2ex}
\includegraphics[scale=0.3]{Bc_BsD_mass.eps}
\hspace{3ex}
\includegraphics[scale=0.3]{Bc_BsD_pot.eps}
\vspace{5ex}
\\
\includegraphics[scale=0.3]{Bc_BDs_mass.eps}
\hspace{4ex}
\includegraphics[scale=0.3]{Bc_BDs_pot.eps}
\caption{In-medium mass (left panel) and mass shift (right panel) of $B_c$ meson (i) only including
the $B^{*}D$ loop (top) and
(ii) only including the $BD^{*}$. $\rho_0 = 0.15  fm^{-3} $ is the symmetric nuclear matter
saturation
density and $\rho_B$ is the baryon density.
\label{Bcmasspot}}
\end{figure}

In Figs.~\ref{Bcmasspot}, we show respectively the mass and mass shift
of $B_c$  meson in symmetric nuclear matter,
varying five different values of the cutoff
($\Lambda$ = 2000, 3000, 4000, 5000, 6000 MeV).
In the figures, $\Delta m$ is the mass shift
($\Delta m = m^{*} - m$) and
$\rho_0 = 0.15$ fm$^{-3}$ is the symmetric nuclear matter saturation
density and $\rho_B$ is the baryon density.

The physical free space masses used in calculation are
$m_{B_{c}} = 6274.5$ MeV, $m_{B^{*}_{c}} = 6333.0$ MeV, and since there
is no experimental data of $B^{*}_c$ we use an average value of
those presented in Ref.~\cite{Martin-Gonzalez:2022qwd}. (See Table~\ref{MesMasTab})

Note that, in Fig.~\ref{Bc_totalmasspot}, $B_{c}$ total mass shift  is provided by
the sum of the two loop contributions as shown in Fig.~\ref{psmeson} (top panel).

\begin{figure}[htb!]
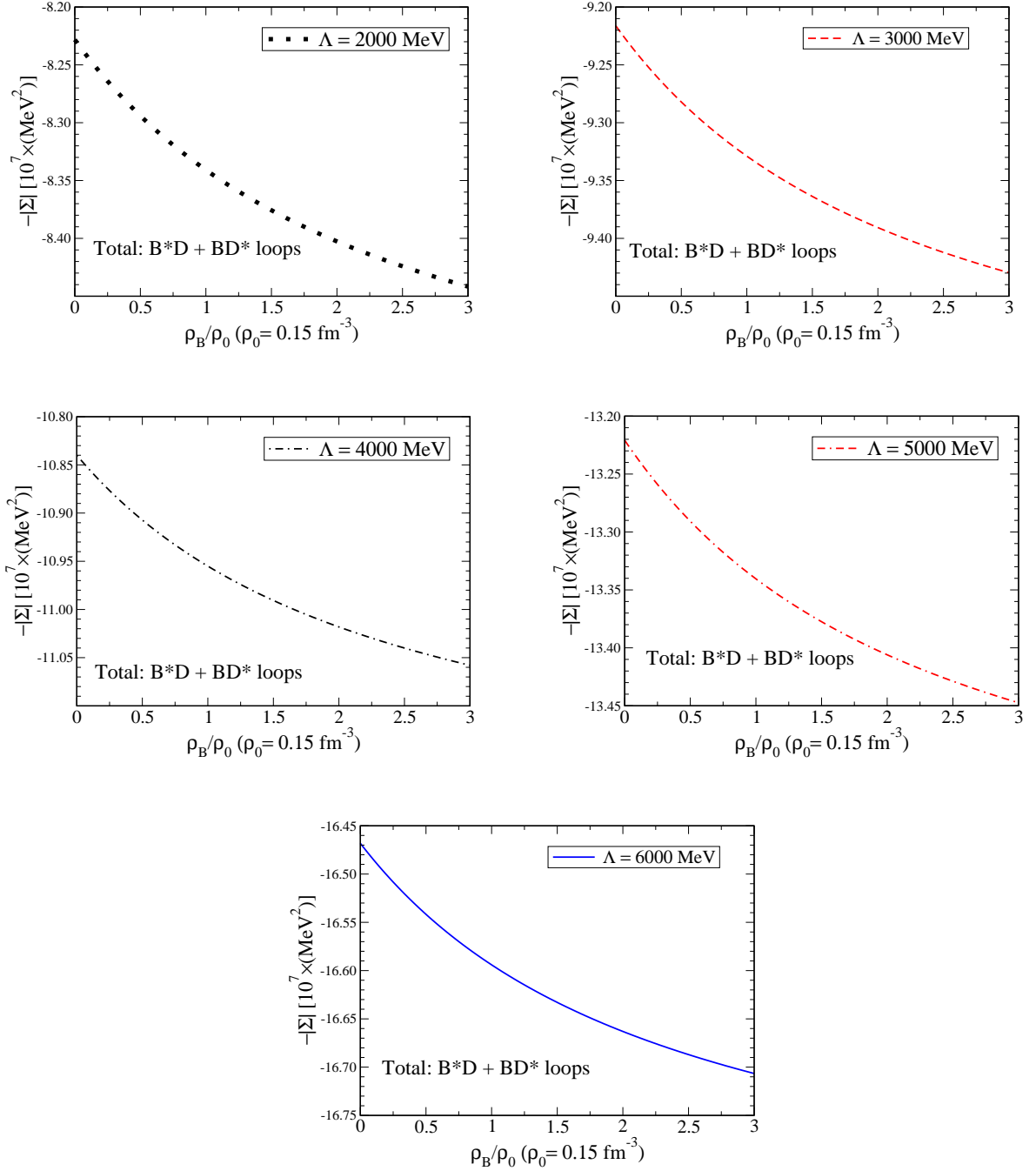

\vspace{3ex}
\includegraphics[scale=0.3]{Bc_SEtotal_L2000RS.eps}
\hspace{5ex}
\includegraphics[scale=0.3]{Bc_SEtotal_L3000RS.eps}
\vspace{5ex}
\\
\includegraphics[scale=0.3]{Bc_SEtotal_L4000RS.eps}
\hspace{5ex}
\includegraphics[scale=0.3]{Bc_SEtotal_L5000RS.eps}
\vspace{5ex}
\\
\centering{
\includegraphics[scale=0.3]{Bc_SEtotal_L6000RS.eps}}
\caption{$B_c$ self-energy values considering total two possible loops. Each figure shows the self-energy for total loop considering both
loops for a different cutoff
value. $\rho_0 = 0.15$  fm$^{-3} $ is the symmetric nuclear matter saturation
density and $\rho_B$ is the baryon density. 
\label{BcSE}}
\end{figure}

In Fig.~\ref{BcSE} it is explicitly shown the density dependence of self-energy
for each loop and its total, besides that, it is shown the density dependence of total 
self-energy for each cutoff value separately for the $B_c$ meson.

These self-energy results show negative shift
in effective masses (or attractive Lorentz scalar potentials),
what can be interpreted as  attractive Lorentz scalar potentials provided by
the effects of the nuclear medium~\cite{Krein:2010vp}.
(Note that, as already commented, in quantum mechanics, the second order perturbation always
gives the negative contribution for the energy, and one can understand
based on this picture when the intermediate state energies are more than that
of the initial and final states.)

The interesting question that arises is whether or not this attraction is strong enough
to bind these mesons to a nucleus.
Thus, the next step should be to look for nucleus bound states.

We remind that, for each case in graphs shown in Fig.~\ref{Bcmasspot}, $m^{0}$ is determined
to reproduce the physical mass $m_{B_c}$ in free space.

As we noted after Eqs.~(\ref{tmass1})-(\ref{tmass3}) as well as
Eq.~(\ref{m0}), one should be careful in comparing each loop contribution
in Table~\ref{BcPotTb}. For each loop contribution
$B^*D, BD^*$ and $B^*D + BD^*$, $m^0$ is fixed as $m^0(B^*D), m^0(BD^*)$
and $m^0(B^*D + BD^*)$ respectively, to reproduce each case
the physical free mass of $m_{B_c}$. 
We can see more details in Table~\ref{Bcm0}.

\begin{table}[htb!]
\caption{ $\Delta m$ for $B_{c}$ meson at $3\rho_{0}$
for all the five cutoff value ($\Lambda$ = 2000, 3000, 4000, 5000, 6000 MeV).
\label{BcPotTb}
}
\begin{center}
\begin{tabular}{c|c|c|c}
\hline
\hline
$\Lambda$ (MeV) &$B^{*}D$ (MeV) &$BD^{*}$ (MeV)
&$B^{*}D+BD^{*}$ (MeV)\\
\hline
\hline
2000		& -125.9 & -172.7 & -172.4 \\
3000      	& -121.4 & -174.6 & -172.5 \\
4000   	  	& -119.6 & -180.4 & -176.1 \\
5000      	& -120.3 & -190.0 & -183.2 \\
6000      	& -122.8 & -202.5 & -192.9 \\
\hline   
\hline
\end{tabular}
\end{center}   
\end{table}

\begin{table}[htb!]
\caption{
$m^0$ considering $B^*D$ loop, $BD^*$ loop and total $B^{*}D + BD^{*}$. 
\label{Bcm0}
}
\begin{center}   
\begin{tabular}{c|c|c|c}
\hline
\hline
$\Lambda$ (MeV) & $m^0$ $(B^{*}D)$ (MeV) & $m^0$ $(BD^{*})$ (MeV)  & $m^0$ $(B^{*}D + BD^{*})$ (MeV) \\
\hline
\hline
2000		& {7906.1}  & {9925.4}   & {11029.6}  \\
3000      	& {8032.6}  & {10313.9}  & {11468.7} \\
4000    	& {8249.0}  & {10913.0}  & {12156.0} \\
5000      	& {8561.5}  & {11732.3}  & {13098.8} \\
6000      	& {8968.6}  & {12766.7}  & {14284.6} \\
\hline   
\hline
\end{tabular}
\end{center}   
\end{table}

By Table~\ref{BcPotTb}, we can conclude
that, for the total mass shift including the $B^{*}D + BD^{*}$ loops,
the $BD^{*}$ loop is more dominant than the $B^{*}D$ loop. This indeed shows that
very naive expectation by comparing the total mass values {\it does not} work.
This may also be caused by the fact that the total mass values
for these two loops are very close.

A possible explanation is as follows. Looking at Eq.~(\ref{BcSE}) for the $B^*D$ loop,
it contains $B^*$ vector meson propagator, which results in the expression Eq.~(\ref{Bsprop}).
For the $BD^*$ loop instead, one may replace $m^*_B \to m^*_D$ in Eq.~(\ref{Bsprop}).
Since $m_{B^*} - m_B \simeq 45 {\rm\,\, MeV} << m_{B^*} - m_{D^*} \simeq 2300 {\rm\,\, MeV}$
and $m_{D^*} - m_D \simeq 140 {\rm\,\, MeV} << m_{B^*} - m_{D^*} \simeq 2300 {\rm\,\, MeV}$,
thus, even in nuclear medium the $B^*$ and $D^*$ role change is expected
to give significant effect of the $D^*$ meson propagator.

Thus, the conclusion is that the $D^*$ meson propagator,
in particular the numerator which represents the spin-1 nature, is much more effectively
contributing to the self-energy integral than that of the $B^*$.

\begin{figure}[htb!]
\vspace{7ex}
\begin{center}
\includegraphics[scale=0.45]{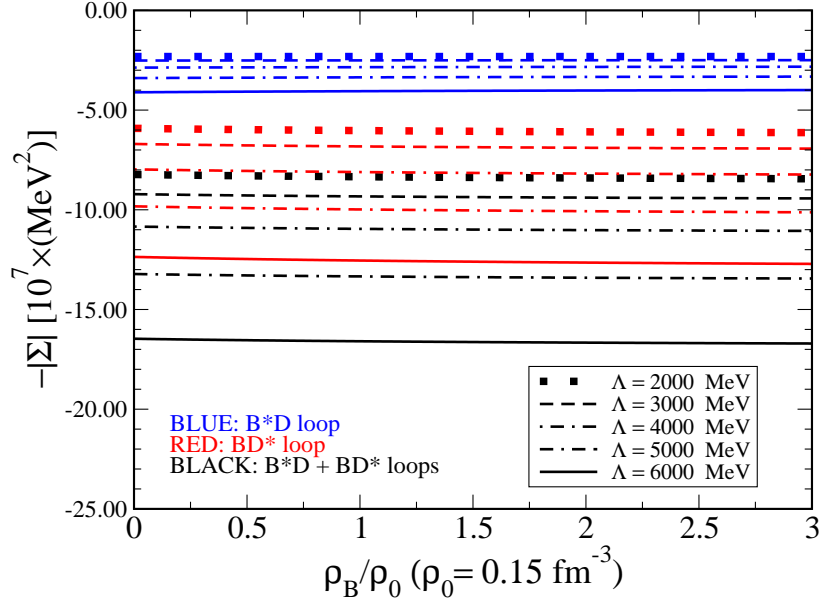}
\caption{Decomposition of the $B_{c}$ meson total self-energy for different cutoff mass values.
\label{BcSEa}}
\end{center}
\end{figure}

\begin{figure}[htb!]
\vspace{7ex}
\hspace{-5ex}
\includegraphics[scale=0.33]{Bc_totalmass.eps}
\hspace{3ex}
\includegraphics[scale=0.33]{Bc_totalpot.eps}
\caption{$B_{c}$ in-medium mass (left) and mas shift (right). $\rho_0 = 0.15$  fm$^{-3} $ is the
symmetric nuclear matter saturation
density and $\rho_B$ is the baryon density.
\label{Bc_totalmasspot}}
\vspace{3ex}
\end{figure}

Note  that in Fig.~\ref{BcSEa} we present the self-energy decomposition of meson $B_c$. It 
seems that the density dependence is small but it is due to the vertical scale.


\begin{figure}[htb!]
\begin{center}
\vspace{6ex}
\includegraphics[scale=0.45]{Bcs_SEtotalRS2.eps}
\caption{$B^{*}_{c}$ self-energies for different cutoff mass values.
\label{BcsSE}}
\end{center}
\end{figure}

\begin{figure}[htb!]
\begin{center}
\vspace{7ex}
\hspace{-6ex}
\includegraphics[scale=0.33]{Bcs_mass.eps}
\hspace{6ex}
\includegraphics[scale=0.33]{Bcspot.eps}
\caption{$B^{*}_{c}$ in-medium mass (left) and mas shift (right).
 $\rho_0 = 0.15$  fm$^{-3} $ is the symmetric nuclear matter saturation
density and $\rho_B$ is the baryon density.
\label{Bcstotalmasspot}}
\end{center}
\end{figure}

\begin{table}[htb!]
\caption{$\Delta m$ in medium for $B^{*}_{c}$ meson at $3\rho_{0}$
for all the five cutoff value ($\Lambda$ = 2000, 3000, 4000, 5000, 6000 MeV).
\label{BcsPotTb}
}
\begin{center} 
\begin{tabular}{c|c}
\hline
\hline
$\Lambda$ (MeV) 	&     $BD$ (MeV)   \\
\hline
\hline
2000		& -29.4 \\
3000      	& -30.1 \\
4000    	& -31.9 \\
5000      	& -34.9 \\
6000      	& -38.8 \\
\hline   
\hline
\end{tabular}
\end{center}  
\end{table}

\begin{table}[htb!]
\caption{
$m^0$ considering $BD$ loop. 
\label{Bcsm0}
}
\begin{center} 
\begin{tabular}{c|c}
\hline
\hline
$\Lambda$  (MeV) 	&  $m^{0}$ $(BD)$  (MeV)    \\
\hline
\hline
2000		& {6449.0} \\
3000      	& {6467.4} \\
4000    	& {6496.8} \\
5000      	& {6539.4} \\
6000      	& {6597.7} \\
\hline   
\hline
\end{tabular}
\end{center}  
\end{table}

Figures~\ref{BcsSE} and~\ref{Bcstotalmasspot} show the decomposition of $B^{*}_c$ meson
self-energy and its mass and mass-shift respectively.

Table~\ref{Bcsm0} shows how $m^{0}$ value varies against the cut-off value, and
Table~\ref{BcsPotTb}
shows the mass shift of $B^{*}_c$ for the each cut-off value.

We also study the in-medium mass shift of the heavy-strange mesons $ B_{s}, B^{*}_{s}, D_{s}$ and
$D^{*}_{s}$
using the same method used for $B_{c}$ and $B^{*}_{c}$.

The free space mass of each meson is,  $m_{B_{s}}=5366.9$ MeV, $ m_{B^{*}_{s}}= 5415.4$ MeV,
$m_{D_{s}}=1968.4 $ MeV and $m_{D^{*}_{s}}= 2112.2$ MeV.

\section{$B^{0}_{s}$ and $B^{*}_{s}$ meson results}
\begin{figure}[htb!]
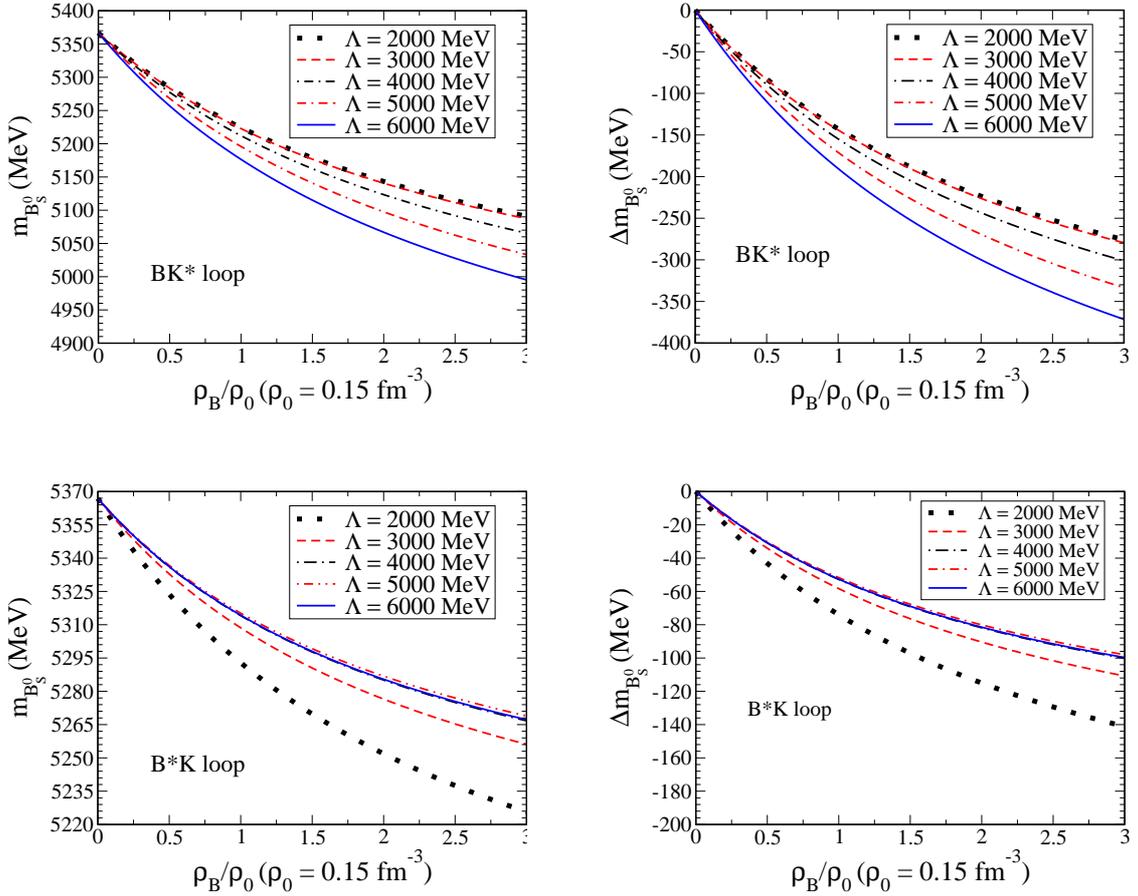

\includegraphics[scale=0.3]{Bs0_BKs_mass.eps}
\hspace{4ex}
\includegraphics[scale=0.3]{Bs0_BKs_pot.eps}
\vspace{5ex}
\\
\includegraphics[scale=0.3]{Bs0_BsK_mass.eps}
\hspace{4ex}
\includegraphics[scale=0.3]{Bs0_BsK_pot.eps}
\caption{In-medium mass (left panel) and mass shift (right panel) of $B^{0}_s$ meson (i) only
including the
$BK^{*}$ loop (top) and (ii) only including the $B^{*}K$. $\rho_0 = 0.15$  fm$^{-3} $ is the
symmetric nuclear matter saturation density and $\rho_B$ is the baryon density.
\label{Bsmasspot}}
\end{figure}

\begin{figure}[htb!]
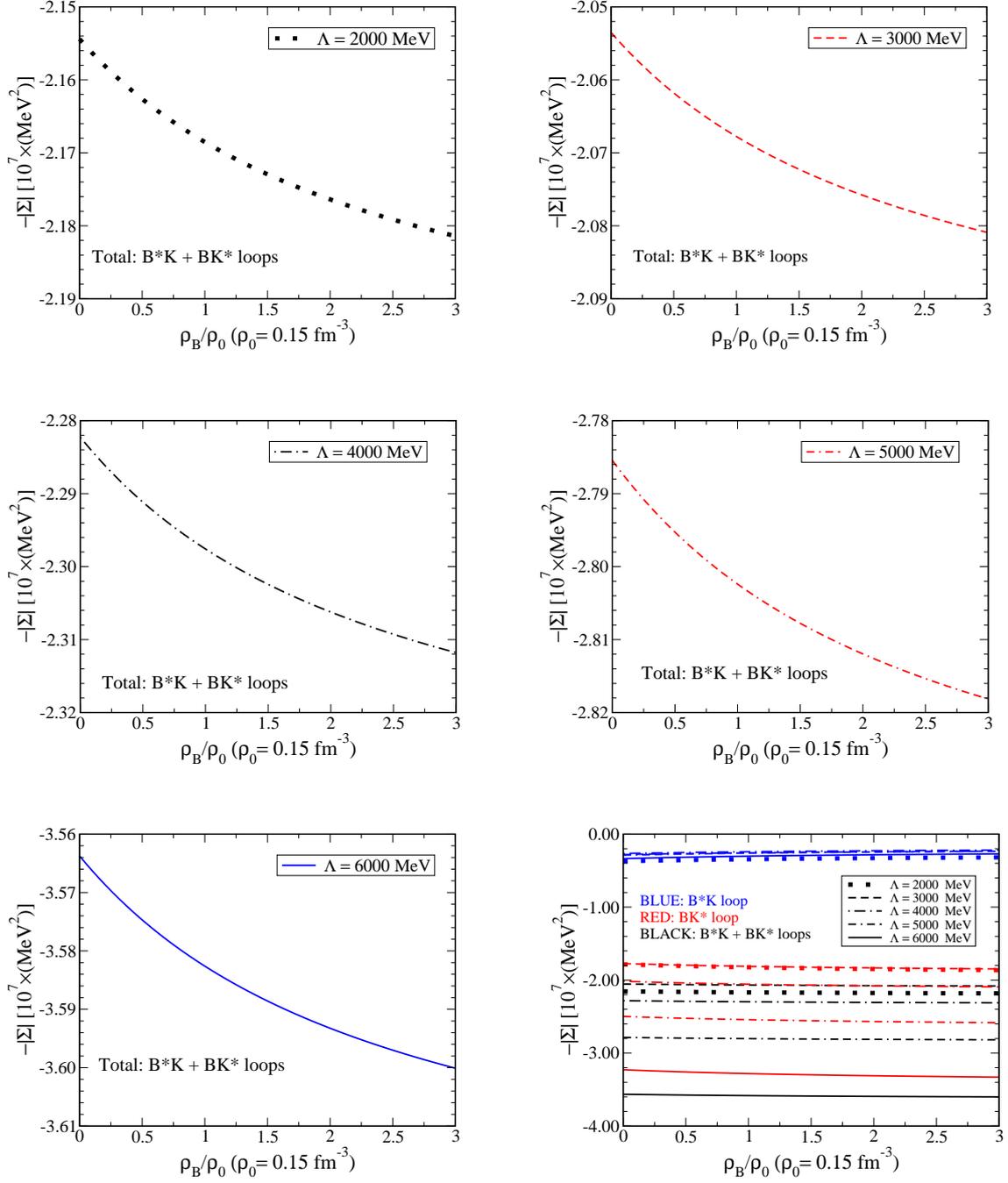

\vspace{3ex}
\includegraphics[scale=0.3]{Bs0_SEtotal_L2000RS.eps}
\hspace{5ex}
\includegraphics[scale=0.3]{Bs0_SEtotal_L3000RS.eps}
\vspace{5ex}
\\
\includegraphics[scale=0.3]{Bs0_SEtotal_L4000RS.eps}
\hspace{5ex}
\includegraphics[scale=0.3]{Bs0_SEtotal_L5000RS.eps}
\vspace{5ex}
\\
\includegraphics[scale=0.3]{Bs0_SEtotal_L6000RS.eps}
\hspace{6ex}
\includegraphics[scale=0.3]{Bs0_SERS.eps}
\caption{$B^{0}_{s}$ self-energy values. Each panel shows
the self-energy for total loop contribution including the
both loops but for a different cutoff value.
The figure in the bottom-right panel shows
a comparison among all of them including 
the total and the two possible loops separately.
\label{Bs0SE}}
\end{figure}

\begin{figure}[htb!]
\begin{center}
\vspace{7ex}
\hspace{-5ex}
\includegraphics[scale=0.33]{Bs0_totalmass.eps}
\hspace{7ex}
\includegraphics[scale=0.33]{Bs0s_totalpot.eps}
\caption{$B^{0}_{s}$ in-medium mass (left) and mas shift (right). $\rho_0 = 0.15$  fm$^{-3} $ is the
symmetric nuclear matter saturation
density and $\rho_B$ is the baryon density.
\label{Bs0totalmasspot}}
\end{center}
\end{figure}

\begin{table}[htb!]
\caption{ $\Delta m$ in medium for $B^{0}_{s}$ meson at $3\rho_{0}$ 
for all the five cutoff value ($\Lambda$ = 2000, 3000, 4000, 5000, 6000 MeV).
\label{Bs0PotTb}
}
\begin{center}   
\begin{tabular}{c|c|c|c}
\hline
\hline
$\Lambda$ (MeV) 	& $B^{*}K$ (MeV)	& $BK^{*}$ (MeV)	& $B^{*}K + BK^{*}$ (MeV)\\
\hline
\hline
2000		& -141.1 & -275.8 & -257.6 \\
3000      	& -110.8 & -279.5 & -261.4 \\
4000     	& -100.2 & -301.2 & -282.5 \\
5000      	& -98.0 & -333.5 & -313.6 \\
6000      	& -99.7 & -371.5 & -350.2 \\
\hline   
\hline
\end{tabular}
\end{center}   
\end{table}

\begin{table}[htb!]
\caption{
$m^0$ considering $B^*K$ loop $BK^*$ loop and total $B^{*}K + BK^{*}$. 
\label{Bs0m0}
}
\begin{center} 
\begin{tabular}{c|c|c|c}
\hline
\hline
$\Lambda$ (MeV)  & $m^0$  $(B^{*}K)$ (MeV) & $m^0$  $(BK^{*})$ (MeV) & $m^0$ $(B^{*}K + BK^{*})$ (MeV) \\
\hline
\hline
2000		& 8114.6  & 14394.5 & 15628.3\\
3000      	& 7536.4  & 14358.3 & 15302.1\\
4000 	  	& 7441.7  & 15180.7 & 16032.1\\
5000      	& 7592.3  & 16688.6 & 17531.3\\
6000      	& 7899.3  & 18750.5 & 19625.9\\
\hline   
\hline
\end{tabular}
\end{center}   
\end{table}

Fig.~\ref{Bsmasspot} shows the mass and the mass shift of $B^{0}_{s}$  where we can check that
$B^{0}_s$ shares a  similarity with $B_c$, since $B^{*}K$ loop gives smaller contribution than
$BK^{*}$ for this  meson total self-energy.
In the same way, a possible explanation is that the
lighter vector meson (excited) $K^{*}$ propagator gives this result.
Figure~\ref{Bs0SE} explicitly shows the self-energy values for each loop and its total.
In Fig.~\ref{Bs0totalmasspot} we show the mass (left) and mass shift (right) of
$B^{0}_s$ considering total loop contribution, and in Table.~\ref{Bs0PotTb} the mass shift values of each loop for each cut-off value at $3\rho_0$.
Table~\ref{Bs0m0} shows the $m_0$ for each cut-off value.\\
%
%

\begin{figure}[htb!]
\vspace{6ex}
\begin{center}
\includegraphics[scale=0.45]{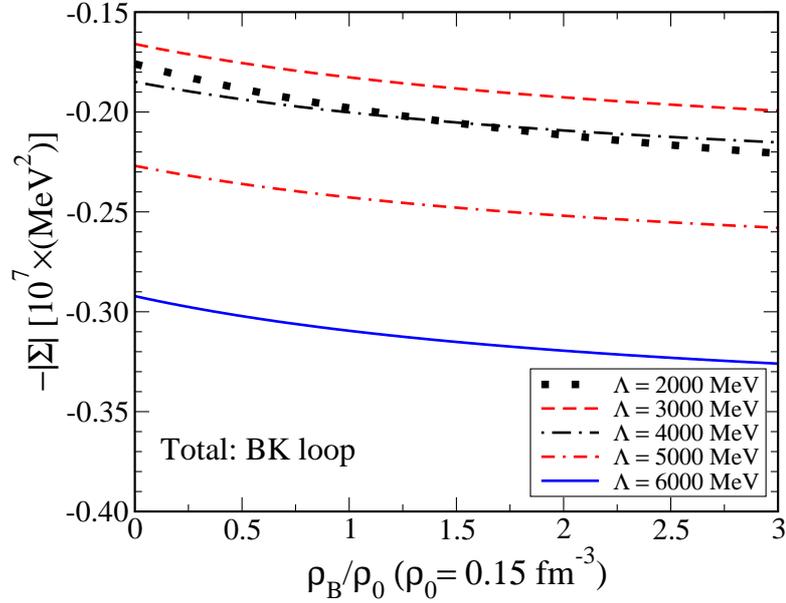}
\caption{$B^{*}_{s}$ self-energies for different cutoff mass values.}
\label{BssSE}
\end{center}
\vspace{5ex}
\end{figure}

\begin{figure}[htb!]
\begin{center}
\vspace{7ex}
\hspace{-10ex}
\includegraphics[scale=0.33]{Bssmass.eps}
\hspace{7ex}
\includegraphics[scale=0.33]{Bsspotential.eps}
\caption{$B^{*}_{s}$ in-medium mass (left) and mass shift (right).
\label{Bssmasspot}}
\end{center}
\end{figure}


\begin{table}[htb!]
\caption{$\Delta m$ in medium for $B^{*}_{s}$ meson at $3\rho_{0}$
for all the five cutoff value ($\Lambda$ = 2000, 3000, 4000, 5000, 6000 MeV).
\label{BcssPotTb}
}
\begin{center}   
\begin{tabular}{c|c}
\hline
\hline
$\Lambda$  	(MeV) &     $BK$    (MeV) \\
\hline
\hline
2000		& -41.6 \\
3000      	& -30.9 \\
4000     	& -28.1 \\
5000      	& -28.7 \\
6000      	& -31.3 \\
\hline   
\hline
\end{tabular}
\end{center}   
\end{table}

\begin{table}[htb!]
\caption{
$m^0$ considering $BK$ loop. 
\label{Bssm0}
}
\begin{center}  
\begin{tabular}{c|c}
\hline
\hline
$\Lambda$ (MeV) 	&    $m^0$ ($BK$) (MeV)     \\
\hline
\hline
2000		& 5575.42 \\
3000      	& 5566.5 \\
4000    	& 5583.5 \\
5000      	& 5621.0 \\
6000      	& 5678.8 \\
\hline   
\hline
\end{tabular}
\end{center}  
\end{table}

Fig.~\ref{Bssmasspot} shows the results obtained for the mass (left), mass shift (right)
and FIg.~\ref{BssSE} shows the  self-energy for each calculated cut-off value
of the vector meson $B^{*}_{s}$ and
Table~\ref{BcssPotTb} gives more details about the mass shift in each cut off value.
Table~\ref{Bssm0} shows $m^0$ for each calculated cut-off value
of the vector meson $B^{*}_{s}$.

\section{$D_{s}$ and $D^{*}_{s}$ meson results}
\begin{figure}[htb!]
\vspace{3ex}
\includegraphics[scale=0.3]{Dcs_DsK_mass.eps}\hspace{3ex}
\hspace{2ex}
\includegraphics[scale=0.3]{Dcs_DsK_pot.eps}
\vspace{5ex}
\\
\includegraphics[scale=0.3]{Dcs_DKs_mass.eps}\hspace{3ex}
\hspace{2ex}
\includegraphics[scale=0.3]{Dcs_DKs_pot.eps}
\caption{In-medium mass (left panel) and mass shift (right panel) of $D_s$ meson (i) only including
the $D^{*}K$ loop (top) and
(ii) only including the $DK^{*}$. $\rho_0 = 0.15$  fm$^{-3} $ is the symmetric nuclear matter
saturation
density and $\rho_B$ is the baryon density.
\label{Dsmasspot}}
\end{figure}

\begin{figure}[htb!]
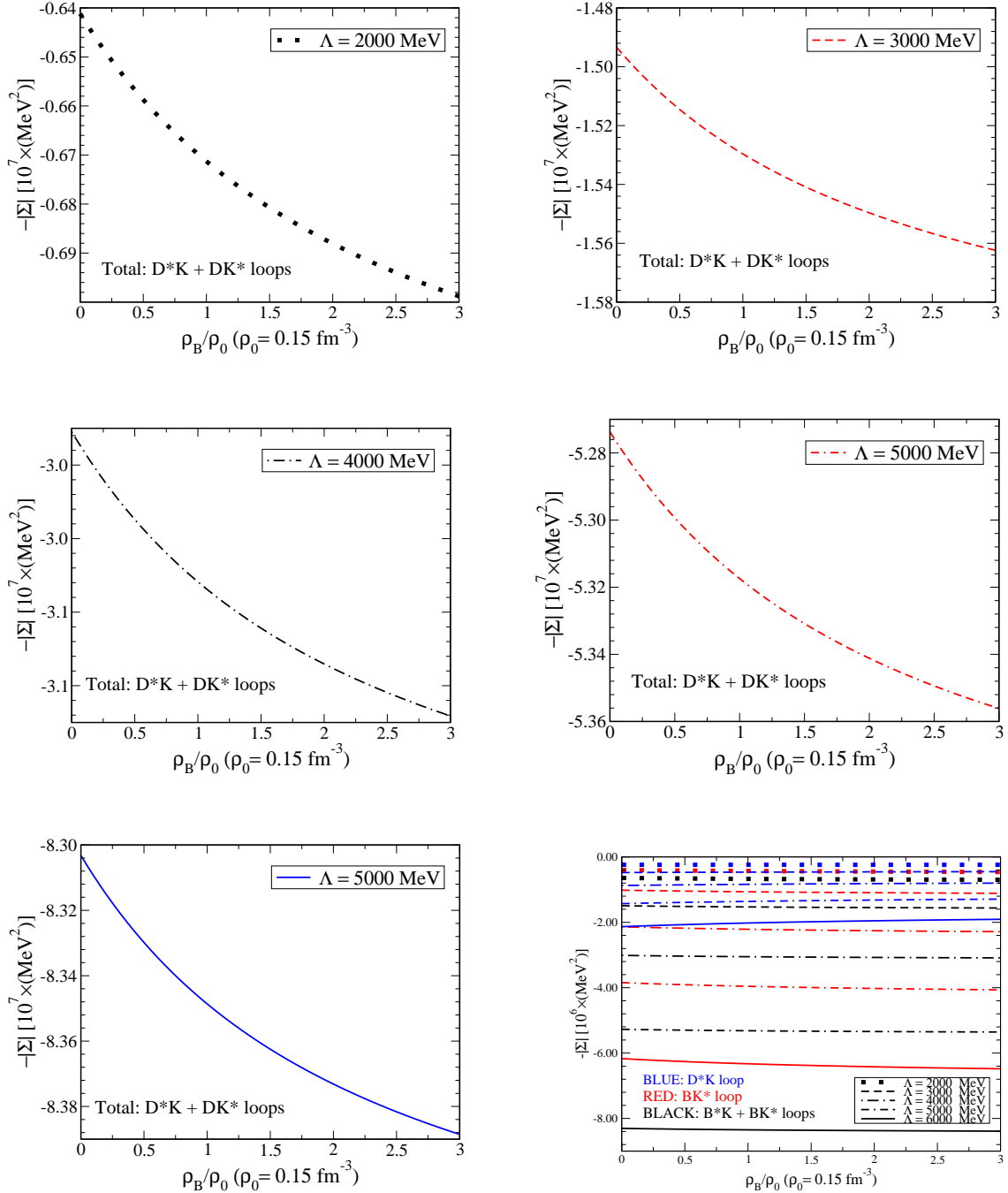

\includegraphics[scale=0.3]{Dcs_SEtotal_L2000RS.eps}
\hspace{5ex}
\includegraphics[scale=0.3]{Dcs_SEtotal_L3000RS.eps}
\vspace{5ex}
\\
\hspace{-2ex}
\includegraphics[scale=0.3]{Dcs_SEtotal_L4000RS.eps}
\hspace{5ex}
\includegraphics[scale=0.3]{Dcs_SEtotal_L5000RS.eps}
\vspace{5ex}
\\
\includegraphics[scale=0.3]{Dcs_SEtotal_L6000RS.eps}
\hspace{7ex}
\includegraphics[scale=0.3]{Dcs_SERS.eps}
\caption{$D_{s}$ self-energy values. Each panel shows the self-energy
for the total loop contribution from both loops but a different cutoff value,
and the figure in the right-bottom panel shows a comparison
among all of them including the total and two different each possible loop.
\label{DcsSE}}
\end{figure}

\begin{figure}[htb!]
\vspace{7ex}
\hspace{-5ex}
\includegraphics[scale=0.33]{Dcs_totalmass.eps}
\hspace{7ex}
\includegraphics[scale=0.33]{Dcs_totalpot.eps}
\caption{$D_{s}$ in-medium mass (left) and mas shift (right). $\rho_0 = 0.15$  fm$^{-3} $ is the
symmetric nuclear matter saturation
density and $\rho_B$ is the baryon density.
\label{Dcs_totalmasspot}}
\vspace{3ex}
\end{figure}

\begin{table}[htb!]
\caption{ $\Delta m$ for $D_{s}$ meson at $3\rho_{0}$
for all the five cutoff value ($\Lambda$ = 2000, 3000, 4000, 5000, 6000 MeV).
\label{DcsPotTb}
}
\begin{center}   
\begin{tabular}{c|c|c|c}
\hline
\hline
$\Lambda$ (MeV)  &$D^{*}K$ (MeV) &$DK^{*}$ (MeV) &  $D^{*}K + DK^{*}$ (MeV)\\
\hline
\hline
2000		& -101.5 & -152.0 & -152.2 \\
3000      	& -115.9 & -194.0 & -183.9 \\
4000   	  	& -126.8 & -224.6 & 207.3 \\
5000   	   	& -131.7 & -243.7 & -221.8 \\
6000      	& -132.5 & -255.1 & -230.3 \\
\hline
\hline
\end{tabular}
\end{center}   
\end{table}
\vspace{5ex}

\begin{table}[htb!]
\caption{
$m^0$ considering $D^*K$ loop $DK^*$ loop and total $D^{*}K + DK^{*}$.
\label{Dcm0}
}
\begin{center}
\begin{tabular}{c|c|c|c}
\hline
\hline
$\Lambda$  (MeV) & $m^0$ ($D^{*}K$) (MeV) & $m^0$ ($DK^{*}$) (MeV)	& $m^0$  ($D^{*}K + DK^{*}$) (MeV)\\
\hline
\hline
2000		& 2503.1& 2809.9& 3207.3\\
3000      	& 2940.5& 3746.6& 4336.8\\
4000  	   	& 3551.1& 5024.7& 5829.5\\
5000      	& 4262.1& 6505.5& 7524.1\\
6000      	& 5021.3& 8097.4& 9322.4\\
\hline
\hline
\end{tabular}
\end{center}
\end{table}


\begin{figure}[htb!]
\vspace{5ex}
\begin{center}
\includegraphics[scale=0.45]{DcssSEtotalRS2.eps}
\caption{$D^{*}_{s}$ self-energies for different cutoff mass values.
\label{DcssSE}}
\end{center}
\end{figure}

\begin{figure}[htb!]
\begin{center}
\vspace{9ex}
\hspace{-10ex}
\includegraphics[scale=0.33]{Dcssmass.eps}
\hspace{7ex}
\includegraphics[scale=0.33]{Dcsspot.eps}
\caption{$D^{*}_{s}$ in-medium mass (left) and mass shift (right).
\label{Dcssmasspot}}
\end{center}
\end{figure}

\begin{table}[htb!]
\caption{$\Delta m$  for $D^{*}_{s}$ meson at $3\rho_{0}$
for all the five cutoff value ($\Lambda$ = 2000, 3000, 4000, 5000, 6000 MeV).
\label{DcssPotTb}
}
\begin{center}   
\begin{tabular}{c|c}
\hline
\hline
$\Lambda$  (MeV)	&     $DK$  (MeV)     \\
\hline
\hline
2000		& -44.6 \\
3000      	& -57.2 \\
4000   	  	& -72.9 \\
5000      	& -88.4 \\
6000      	& -102.4 \\
\hline
\hline
\end{tabular}
\end{center}   
\end{table}

\begin{table}[htb!]
\caption{
$m^0$ considering $DK$ loop.
\label{Dcsm0}
}
\begin{center}
\begin{tabular}{c|c}
\hline
\hline
$\Lambda$ (MeV) 	&   $m^0$ ($DK$) (MeV)    \\
\hline
\hline
2000		& 2198.66 \\
3000      	& 2288.6 \\
4000  	   	& 2441.2 \\
5000      	& 2656.0 \\
6000      	& 2924.4 \\
\hline
\hline
\end{tabular}
\end{center}
\end{table}

Figures~\ref{DcssSE} and~\ref{Dcssmasspot} show the results for the self-energy
the in-medium mass (left) and mass shift (right)
of the vector meson $D^{*}_{s}$.
Table~\ref{DcssPotTb} shows a detailed analyses of its mass shift value for each cut-off value.
Table~\ref{Dcsm0} shows $m_0$ value for each cut-off value.

After presenting all the obtained results we summarize in
Table~\ref{NewMesMasTab} the density dependence of
the in-medium mass (where $\rho_0 = 0.15$ fm$^{-3}$ is the nuclear matter saturation density
and $\rho_B$ is the baryon density). We have  used the same parameters as those previously
used in Ref.~\cite{Tsushima:2020gun} but now we include the values obtained in this work.
(For $m_{B_c^{*}}$ at $\rho_B=0$ value we extract the average value from \cite{Martin-Gonzalez:2022qwd}.)

\begin{table}[htb!]
\caption{
Density dependence of meson (effective) masses in MeV calculate in the QMC
model with $m_{u,d}=5$ MeV, $m_s=250$ MeV, $m_c=1270$ MeV and
$m_b=4200$ MeV. ($\rho_0 = 0.15$ fm$^{-3}$ below.)
\label{NewMesMasTab}
}
\begin{center}
\begin{tabular}{c|c|c|c|c}
\hline
\hline
&$\rho_B=0$ (MeV) &$\rho_B=\rho_0$ (MeV)
&$\rho_B=2\rho_0$ (MeV) &$\rho_B=3\rho_0$ (MeV)\\
\hline
\hline
$m_K$       &493.7  &430.5  &393.6  &369.0  \\
$m_{K^*}$   &893.9  &831.9  &797.2  &775.0  \\
$m_D$       &1867.2 &1805.2 &1770.6 &1748.4 \\
$m_{D^*}$   &2008.6 &1946.9 &1912.9 &1891.2 \\
$m_B$       &5279.3 &5218.2 &5185.1 &5164.4 \\
$m_{B^*}$   &5324.7 &5263.7 &5230.7 &5210.2 \\
$m_{B^{+}_{c}}$  & 6274.5  & 6184.1 & 6134.0 & 6102.1  \\
$m_{B^{*}_{c}}$   &6333.0  &6318.5  &6309.6 &6303.6 \\
$m_{B^{0}_{s}}$   &5366.9  &5233.8  &5158.2 &5109.3  \\
$m_{B^{*}_{s}}$   &5415.4  &5394.9  &5382.4  &5373.8  \\
$m_{D^{\pm}_{s}}$ &1968.4  &1890.4  &1845.5  &1816.2 \\
$m_{D^{*\pm}_{s}}$ &2112.2  &2091.8  &2077.8  &2067.5  \\
\hline
\hline
\end{tabular}
\end{center}
\end{table}

\pagebreak

\chapter{Comparison with heavy quarkonia}

As we previously mentioned, some of our interest in studying 
mesons $B_c$  and $B_c^*$ relies on the fact that they are composed of
a bottom and a charm quarks. Then, we can very naively expect
that they may present the properties and characteristics between the
mesons composed of only by a pair of charm-anticharm and a pair of bottom-antibottom.

\begin{figure}[htb!]
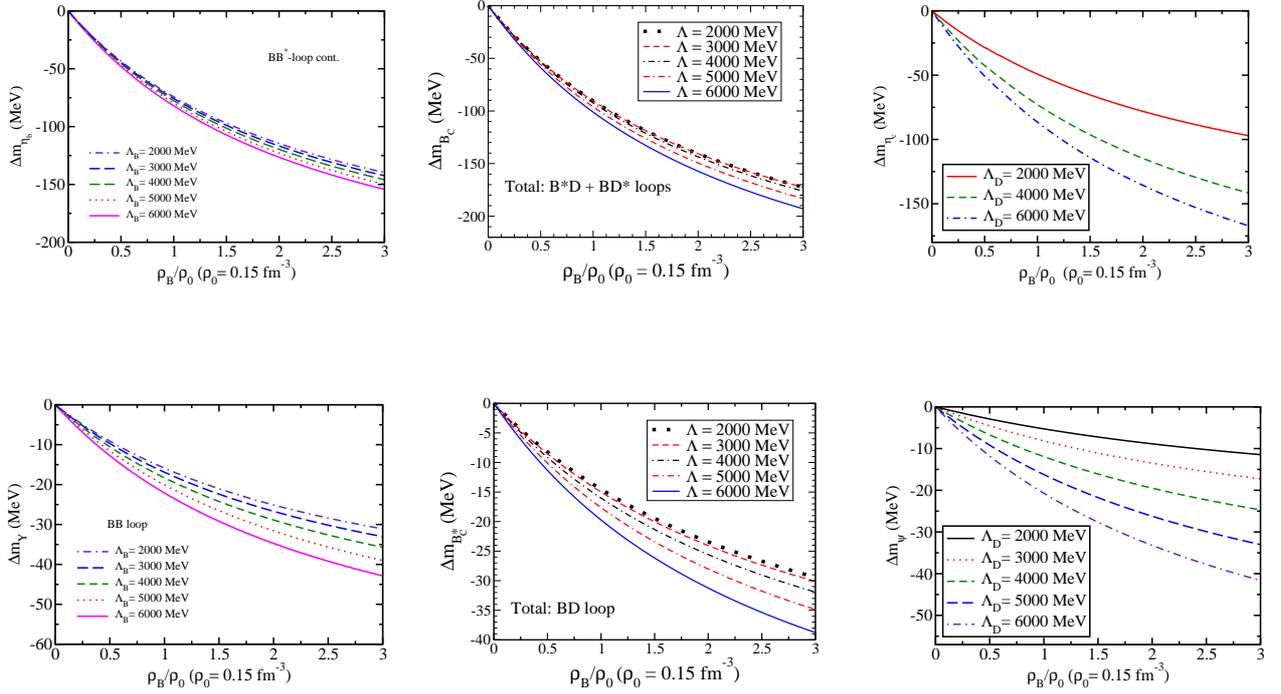

\begin{center}
\hspace{-6ex}
\vspace{5ex}
\includegraphics[width=5cm]{etab_BBs.eps}\hspace{2ex}
\includegraphics[width=5cm]{Bc_totalpot.eps}\hspace{4ex}
\includegraphics[width=5cm]{Dmetac_DDsOnly.eps}
\vspace{3ex}
\\
\hspace{-6ex}
\includegraphics[width=5cm]{Upsilon_pot_BB.eps}\hspace{3ex}
\includegraphics[width=5cm]{Bcspot.eps}\hspace{4ex}
\includegraphics[width=5cm]{DJPsi_in_medium_XiOff_OnlyDDbar.eps}
\caption{Comparison of the in medium mass shift of pseudoscalar (top) and vector (bottom) mesons,
$\eta_b$ (top-left), $B_c$ (top-center) $\eta_c$ (top-right)
$\Upsilon$ (bottom-left), $B^{*}_c$ (bottom-center) and  $J/\Psi$ (bottom-right). 
$\rho_0 = 0.15$  fm$^{-3} $ is the symmetric nuclear matter saturation
density and $\rho_B$ is the baryon density.
\label{BcspotCompar}}
\end{center}
\end{figure}

We have found that $B^{*}_c$ mass shift lies
in-between of the $\Upsilon$ and $J/\psi$ properties, however,
$B_c$ does not present such an expected simple characteristic.
A more detailed comparison is shown in Fig.~\ref{BcspotCompar}.

The $B_c$ mass shift, however, presents a more negative 
shift than $\eta_c$ and $\eta_b$ when including the $B^*D +
BD^*$ (fig.~\ref{BcspotCompar}). The same happens
by including the $BD^*$ loop, a less
negative shift occur when including the $B^*D$ loop, though
(not explicitly shown in Fig.~\ref{BcspotCompar}).

In Fig.~\ref{BcspotCompar}, $\eta_b$ self-energy includes
only the $BB^{*}$ loop contribution
and its mass shift range from -75 to -82 MeV, for the cutoff $\Lambda$ values of
2000, 3000, 4000, 5000 and 6000 MeV at $\rho_0$.
For the $\eta_c$ self-energy, it only includes the
$DD^{*}$ loop contribution and corresponds to the mass
shift value range from -41.18 to -86.49 MeV,
for the cutoff $\Lambda_D$ values of 2000, 4000 and 6000  at $\rho_0$.


\chapter{Conclusion}

In this thesis, extending the previous works
of $J/\psi, \eta_c, \eta_b$ and $\Upsilon$ mass shift
in symmetric nuclear
matter ~\cite{Krein:2010vp,Zeminiani:2021xvw,Zeminiani:2021vaq,Zeminiani:2020aho},
we have estimated, for the first time, the
$B_{c}, B^{*}_{c}, B_{s}, B^{*}_{s}, D_{s}$ and $D^{*}_{s}$  
meson mass shift in symmetric nuclear matter.
We have estimated the meson self-energy loop contributions using an effective flavor SU(5) symmetry based Lagrangians with one universal coupling constant
to be able to relate various
coupling constants appearing in the effective Lagrangian density for each meson.
The value for the universal coupling constant is calculated from the experimental
data with the vector meson dominance (VMD) model, namely by the data $\Upsilon \to e^+ e^-$.
The in medium masses of mesons $B, B^{*}, D, D^{*}, K, 
K^{*}$ were calculated using the quark-meson coupling model. 

For the studies of $\Upsilon$ and $\eta_{b}$ in-medium mass
shift~\cite{Zeminiani:2021vaq},
we have used a similar approach using an effective flavor
SU(5) symmetry based Lagrangian, and the anomalous coupling 
one, but in this work we have not included the latter.

For the $B_{c}$ pseudoscalar meson, instead,
we have to include two possible loops $BD^{*}$ and $B^{*}D$
for the self-energy processes,
and no anomalous coupling loops.
Note that, these two loop contributions are based on the same footing
as those for the $\eta_c$ and $\eta_b$ (PPV: pseudoscalar-pseudoscalar-vector)
coupling from the same SU(5) symmetric Lagrangian.

We have followed the same procedure for the pseudoscalar 
meson $B_{s}$, considering only the first order loops
$BK^{*}$ and $B^{*}K$, and for $D_{s}$, including
the first order, the loops $DK^{*}$ and $D^{*}K$.

Since $B^{*}_{c}$ is a vector meson and our effective Lagrangians
are obtained by the PPV interaction, the intermediate states is composed
by $BD$ loop, and, for this same reason $B^{*}_{s}$ and
$D^{*}_{s}$ the included loops for each of them
are respectively $BK$ and $DK$.

For each meson we have presented results for the in medium masses,
mass shift (scalar potential) and self-energy, varying the cutoffs 
($\Lambda$ = 2000, 3000, 4000, 5000, 6000 MeV), following 
our previous works and we have used a dipole form factor.
  
Our results show that $BD^{*}$ loop self-energy
gives a larger contribution than that of the $B^{*}D$,
and this disagrees with our initial very naive expectation based on
non-relativistic quantum mechanics second-order perturbation theory.
It implies that the lighter vector meson $D^{*}$ than $B^*$
in the self-energy loop gives more effective contribution due to the vector
meson propagator Lorentz structure.

We have found that the $B_c$ meson mass shift at $3 \rho_0$
varies from $-152.2$ MeV ($\Lambda=$2000 MeV) 
to $-230.3$ MeV ($\Lambda=$ 6000 MeV). 

While the results of mass shift for $B^{*}_{c}$ varies from
$-29.4$ MeV ($\Lambda=$2000 MeV) 
to $-38.8$ MeV ($\Lambda=$6000 MeV).

Our results can be
summarized as $|\Delta m_{\eta_c}| < |\Delta m_{\eta_b}| < |\Delta m_{B_c}|$
and $|\Delta m_{J/\Psi}| < |\Delta m_{B^*_c}| < |\Delta m_{\Upsilon}|$.

Besides that, we have also studied and presented
the in-medium masses, mass shift and their density dependence
for the mesons, $B_s, B_s^*, D_s$ and $D_s^*$.

Based on the studies made in this thesis, a few future extended studies
are planned: (i) Study of meson-nucleus bound states for some selected
mesons in $B_c, B_c^*, B_s, B_s^*, D_s$ and $D_s^*$,
(ii) study of the in-medium properties of these mesons, such as
weak-interaction properties, electromagnetic properties,
and (iii) reactions involving these mesons in (heavy mass) nucleus.

\chapter{Appendix A}

\section{Poles in integration}

\begin{figure}[htb!]
\includegraphics[scale=0.5]{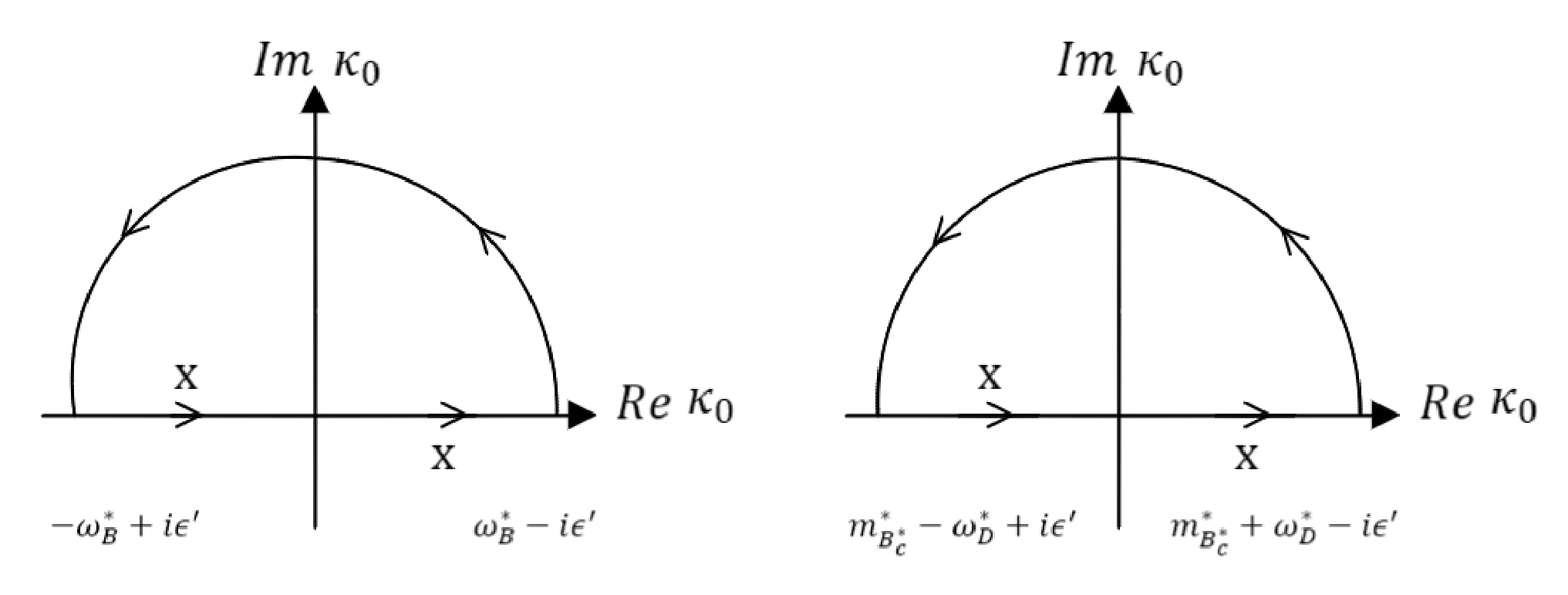}
\caption{Integration of $\kappa_{0}$ in the complex plane.
\label{poleint}}
\end{figure}

In this appendix we show how we perform the complex
integration of $\kappa_{0}$
 in Eqs.~(\ref{BcInt}) and~(\ref{BcsInt}) by showing the corresponding
figures in Fig.~\ref{poleint}, left and right pannels, respectively.

For Eqs.~(\ref{BcInt}) and~(\ref{BcsInt})
the integration contains poles in
$\kappa^{2}_{0} = {\vec{\kappa^{2}}+m^{2}}$,
and this can be handled by the standard Cauchy integral.

With $\kappa_{0} + V \ra \kappa_{0}$ denoting the vector potential shifted,
or changing the integral variables.
By going into the complex plane considering $\kappa_{0}$ as a complex variable
we can take the paths shown
in~\ref{poleint}.

Our choice should be the one that the integral under the circular area goes to zero when 
the radius R goes to infinite.


\text For the left hand side diagram in Fig~\ref{poleint} we proceed,

\begin{equation}
\kappa^{2}-m^{*2}_{B}+i\epsilon = k^{2}_{0} -
\left(\vec \kappa^{2} + m^{*2}_{B} \right) +i\epsilon = 0,
\nonumber
\end{equation}

\begin{equation}
(\vec \kappa^{2} + m^{*2}_{B}) \equiv \omega^{*2}_{B},
\nonumber
\end{equation}

\begin{equation}
\omega^{*}_{B} = \sqrt{\vec \kappa^{2}+m^{*2}_{B}},
\nonumber
\end{equation}

\begin{equation}
\begin{aligned}
\kappa_{0}&= \pm \sqrt{\omega^{*2}_{B} -i\epsilon},\\
&= \pm \omega^{*}_{B} \mp i\epsilon,
\end{aligned}
\nonumber
\end{equation}

\begin{equation}
k_{0} = - \omega^{*}_{B}.
\label{ko}
\end{equation}


\text Similarly, for the right hand side diagram in Fig~\ref{poleint} we proceed,

\begin{equation}
\begin{aligned}
(p-\kappa)^{2}-m^{*2}_{D}+ i\epsilon &= (m^{*}_{B^{*}_{c}}-\kappa_{0})^{2}-(\vec \kappa^{2}+m^{*2}
_{D})+i\epsilon
= 0,\\
&\therefore (\kappa_{0}-m^{*}_{B^{*}_{c}})^{2} = (\vec \kappa + m^{*2}_{D})^{2} - i\epsilon,
\end{aligned}
\nonumber
\end{equation}

\begin{equation}
\begin{aligned}
\kappa_{0}-m^{*}_{B^{*}_{c}} &= \pm \sqrt{\omega^{*2}_{D} -i\epsilon},\\
					&= \pm \omega^{*}_{D} \mp i\epsilon,\\
\end{aligned}
\nonumber
\end{equation}

\begin{equation}
 \kappa_{0}= m^{*}_{B^{*}_{c}}-\omega^{*}_{D}.
 \label{ko2}
\end{equation}


The relations in Eq.~(\ref{ko}) and~(\ref{ko2}) are used to define $\kappa_{0}$ poles in the
complex plane.

\chapter{AppendixB}

\section{SU(5) generators}

Here we present the $\lambda_i$ matrices which are the 
SU(5) generators in the fundamental representation of
\cite{Stancu:2022xax}.

Below we explicitly show the SU(5) ''Gell-Mann'' matrices used,
and getting the physical $P$ and $V$ meson matrices of Eqs. (2.10) and (2.11) respectively.
The first 15 matrices are an extension of the SU(4) generators 
with one 0-row and one 0-column added.

\begin{eqnarray}
\lambda_1   &  = & 
\left( 
\begin{array}{ccccc}
0 & 1 & 0 & 0 & 0\\
1 & 0 & 0 & 0 & 0\\ 
0 & 0 & 0 & 0 & 0 \\
0 & 0 & 0 & 0 & 0 \\
0 & 0 & 0 & 0 & 0 \end{array} 
\right)
,\,\,\,
\lambda_2 = 
\left(
\begin{array}{ccccc}
0 & - i & 0 & 0 & 0 \\
i & 0 & 0 & 0 & 0 \\ 
0 & 0 & 0 & 0 & 0 \\
0 & 0 & 0 & 0 & 0 \\
0 & 0 & 0 & 0 & 0 \end{array}
\right)
,\,\,\,
\lambda_3   =  
\left(
\begin{array}{ccccc}
1 & 0 & 0 & 0 & 0 \\
0 & - 1 & 0 & 0 & 0 \\ 
0 & 0 & 0 & 0 & 0 \\
0 & 0 & 0 & 0 & 0 \\
0 & 0 & 0 & 0 & 0 \end{array}
\right),
\nonumber
\end{eqnarray} 
\begin{eqnarray}
\lambda_4 = 
\left(
\begin{array}{ccccc}
0 & 0 & 1 & 0 & 0 \\
0 & 0 & 0 & 0 & 0 \\ 
1 & 0 & 0 & 0 & 0 \\
0 & 0 & 0 & 0 & 0 \\
0 & 0 & 0 & 0 & 0 \end{array}
\right)
,\,\,\,
\lambda_5 =  
\left(
\begin{array}{ccccc}
0 & 0 & - i & 0 & 0 \\
0 & 0 & 0 & 0 & 0 \\ 
i & 0 & 0 & 0 & 0 \\
0 & 0 & 0 & 0 & 0 \\
0 & 0 & 0 & 0 & 0 \end{array}
\right) 
,\,\,\,
\lambda_6 = 
\left(
\begin{array}{ccccc}
0 & 0 & 0 & 0 & 0 \\
0 & 0 & 1 & 0 & 0 \\ 
0 & 1 & 0 & 0 & 0 \\
0 & 0 & 0 & 0 & 0 \\
0 & 0 & 0 & 0 & 0 \end{array}
\right), 
\nonumber
\end{eqnarray}
\begin{eqnarray}
\lambda_7 = 
\left(
\begin{array}{ccccc}
0 & 0 & 0 & 0 & 0 \\
0 & 0 & - i & 0 & 0 \\ 
0 & i & 0 & 0 & 0 \\
0 & 0 & 0 & 0 & 0 \\
0 & 0 & 0 & 0 & 0 \end{array}
\right) 
,\,\,\,
\lambda_8    =
\frac{1}{\sqrt{3}}
\left(
\begin{array}{ccccc}
1 & 0 & 0 & 0 & 0 \\
0 & 1 & 0 & 0 & 0 \\ 
0 & 0 & - 2 & 0 & 0 \\
0 & 0 & 0 & 0 & 0 \\ 
0 & 0 & 0 & 0 & 0 \end{array}
\right) 
,\,\,\,
\lambda_9 = 
\left(
\begin{array}{ccccc}
0 & 0 & 0 & 1 & 0 \\
0 & 0 & 0 & 0 & 0 \\ 
0 & 0 & 0 & 0 & 0 \\
1 & 0 & 0 & 0 & 0 \\
0 & 0 & 0 & 0 & 0 \end{array}
\right), 
\nonumber
\end{eqnarray}
\begin{eqnarray}
\lambda_{10}    = 
\left(
\begin{array}{ccccc}
0 & 0 & 0 & - i & 0\\
0 & 0 & 0 & 0 & 0 \\ 
0 & 0 & 0 & 0 & 0 \\
i & 0 & 0 & 0 & 0 \\
0 & 0 & 0 & 0 & 0 \end{array}
\right) 
,\,\,\,
\lambda_{11} = 
\left(
\begin{array}{ccccc}
0 & 0 & 0 & 0 & 0 \\
0 & 0 & 0 & 1 & 0\\ 
0 & 0 & 0 & 0 & 0 \\
0 & 1 & 0 & 0 & 0 \\
0 & 0 & 0 & 0 & 0 \end{array}
\right) 
,\,\,\,
\lambda_{12} = 
\left(
\begin{array}{ccccc}
0 & 0 & 0 & 0 & 0 \\
0 & 0 & 0 & - i & 0 \\ 
0 & 0 & 0 & 0 & 0 \\
0 & i & 0 & 0 & 0 \\
0 & 0 & 0 & 0 & 0 \end{array}
\right), 
\nonumber
\end{eqnarray}
\begin{eqnarray}
\lambda_{13}  &  = & 
\left(
\begin{array}{ccccc}
0 & 0 & 0 & 0 & 0 \\
0 & 0 & 0 & 0 & 0 \\ 
0 & 0 & 0 & 1 & 0 \\
0 & 0 & 1 & 0 & 0 \\
0 & 0 & 0 & 0 & 0 \end{array}
\right) 
,\,\,\,
\lambda_{14} = 
\left(
\begin{array}{ccccc}
0 & 0 & 0 & 0 & 0 \\
0 & 0 & 0 & 0 & 0 \\ 
0 & 0 & 0 & - i & 0 \\
0 & 0 & i & 0 & 0 \\
0 & 0 & 0 & 0 & 0 \end{array}
\right) 
,\,\,\,
\lambda_{15} = 
\frac{1}{\sqrt{6}}
\left(
\begin{array}{ccccc}
1 & 0 & 0 & 0 & 0 \\
0 & 1 & 0 & 0 & 0 \\ 
0 & 0 & 1 & 0 & 0 \\
0 & 0 & 0 & - 3 & 0 \\
0 & 0 & 0 & 0 & 0 \end{array}
\right). 
\nonumber
\end{eqnarray}
The additional matrices are
\begin{eqnarray}
\lambda_{16} = 
\left(
\begin{array}{ccccc}
0 & 0 & 0 & 0 & 1 \\
0 & 0 & 0 & 0 & 0 \\ 
0 & 0 & 0 & 0 & 0 \\
0 & 0 & 0 & 0 & 0 \\
1 & 0 & 0 & 0 & 0 \end{array}
\right) 
,\,\,\,
\lambda_{17} = 
\left(
\begin{array}{ccccc}
0 & 0 & 0 & 0 & - i \\
0 & 0 & 0 & 0 & 0 \\ 
0 & 0 & 0 & 0 & 0 \\
0 & 0 & 0 & 0 & 0 \\
i & 0 & 0 & 0 & 0 \end{array}
\right) 
\lambda_{18} = 
\left(
\begin{array}{ccccc}
0 & 0 & 0 & 0 & 0 \\
0 & 0 & 0 & 0 & 1 \\ 
0 & 0 & 0 & 0 & 0 \\
0 & 0 & 0 & 0 & 0 \\
0 & 1 & 0 & 0 & 0 \end{array}
\right), 
\nonumber
\end{eqnarray}
\begin{eqnarray}
\lambda_{19} = 
\left(
\begin{array}{ccccc}
0 & 0 & 0 & 0 & 0 \\
0 & 0 & 0 & 0 & - i \\ 
0 & 0 & 0 & 0 & 0 \\
0 & 0 & 0 & 0 & 0 \\
0 & i & 0 & 0 & 0 \end{array}
\right) 
,\,\,\,
\lambda_{20} = 
\left(
\begin{array}{ccccc}
0 & 0 & 0 & 0 & 0 \\
0 & 0 & 0 & 0 & 0 \\ 
0 & 0 & 0 & 0 & 1 \\
0 & 0 & 0 & 0 & 0 \\
0 & 0 & 1 & 0 & 0 \end{array}
\right) 
,\,\,\,
\lambda_{21} = 
\left(
\begin{array}{ccccc}
0 & 0 & 0 & 0 & 0 \\
0 & 0 & 0 & 0 & 0 \\ 
0 & 0 & 0 & 0 & - i \\
0 & 0 & 0 & 0 & 0 \\
0 & 0 & i & 0 & 0 \end{array}
\right), 
\nonumber
\end{eqnarray}
\begin{eqnarray}
\lambda_{22}  &  = & 
\left(
\begin{array}{ccccc}
0 & 0 & 0 & 0 & 0 \\
0 & 0 & 0 & 0 & 0 \\ 
0 & 0 & 0 & 0 & 0 \\
0 & 0 & 0 & 0 & 1 \\
0 & 0 & 0 & 1 & 0 \end{array}
\right) 
,\,\,\,
\lambda_{23} = 
\left(
\begin{array}{ccccc}
0 & 0 & 0 & 0 & 0 \\
0 & 0 & 0 & 0 & 0 \\ 
0 & 0 & 0 & 0 & 0 \\
0 & 0 & 0 & 0 & - i \\
0 & 0 & 0 & i & 0 \end{array}
\right) 
,\,\,\,
\lambda_{24} = 
\frac{1}{\sqrt{10}}
\left(
\begin{array}{ccccc}
1 & 0 & 0 & 0 & 0 \\
0 & 1 & 0 & 0 & 0 \\ 
0 & 0 & 1 & 0 & 0 \\
0 & 0 & 0 & 1 & 0 \\
0 & 0 & 0 & 0 & - 4 \end{array}
\right). 
\nonumber
\label{su5gen}
\end{eqnarray}
\\
\\

\section{Adjoint Representation \label{Appadjoint}}

To obtain the matrix representations (notation) of
the pseudo-scalar and the vector mesons
(irreducible tensors) {$P^a_b$} and $V_{\mu b}^{a}$,
we use the adjoint representation.

The matrix representation of the generator $(T_a)_{bc}$ is given by the adjoint
representation $(\tilde{T}_a)_{bc}$,
\begin{equation}
({T}_a)_{bc} = -if_{abc} \equiv [ad(T_a)]_{bc} \equiv (\tilde{T}_a)_{bc},
\end{equation}
They satisfy, $[T_a,T_b] = i f_{abc} T_c = [\tilde{T}_a,\tilde{T}_b] = i f_{abc}\tilde{T}_c$.
Using the Cartesian components $\phi$,

\begin{eqnarray}
\phi = 
\left(
\begin{array}{c}
\phi_1 \\
\vdots \\
\phi_n
\end{array}\right)
= (\phi)_a.
\end{eqnarray}

Then, the vector notation $\phi_b$ in the adjoint representation,
or, the $\phi = T_b\phi_b$ transforms under the
general SU(N) group transformation,
\begin{eqnarray}
T_{b}\phi_{b} \ra T_{b}\phi_{b}^{'} & =& \nonumber
T_b [exp(-i\theta^{a}\tilde{T}_{a})]_{b}^{c}\phi_{c},\\ \nonumber
& \simeq &T_{b}[1 - i\theta^{a}\tilde{T}_{a}]_{b}^{c}\phi_{c},\\ \nonumber
& =&T_{b}\phi_{b} -iT_{b}\theta^{a}(\tilde{T}_{a})_{b}^{c}\phi_{c},\\ \nonumber
& =&T_{b}\phi_{b} -iT_{b}\theta^{a}(-if_{abc})\phi_{c},\\ \nonumber
& =&T_{b}\phi_{b} -iT_{b}\theta^{a}(if_{acb})\phi_{c},\\ \nonumber
& =&T_{b}[\phi_{b} -i\theta^{a}(if_{acb})T_{b}\phi_{c}],  \hspace{25ex}
(b \leftrightarrow c,  \,\, {\rm next\,\, line}),\\ \nonumber
& =&T_{b}\phi_{b} -i\theta^{a}(if_{abc})T_{c}\phi_{b},\\
& =&\phi -i\theta^{a}(if_{abc})T_{c}\phi_{b}.
\label{vecrep}
\end{eqnarray}

In the same way, in the matrix notation of $\phi$ in the adjoint representation
is given by,
\begin{eqnarray}
(\phi)^{b}_{a} = \phi_{B}(T_{B})^{b}_{a} = \phi^{B}(T_{B})^{b}_{a}
\hspace{5ex}(B = 1, 2, ..., N).
\label{c}
\end{eqnarray}
In the above, $T_B$ is the generator. We define the unitary operator $U$ as:
\begin{eqnarray}
U \equiv exp(-i\theta^{A}T^{A}) = exp (-i\theta^{A}T_{A}),
\end{eqnarray}
and the $\phi$ transformations as $\phi^{'}$, by changing notation as
$(A,B,C) \to (a,b,c)$:
\begin{eqnarray}
(\phi) \ra \phi^{'}& =& U\phi U^{\dagger},\nonumber \\
& =&U\phi_{b}T_{b}U^{\dagger},\nonumber \\
& \simeq& \phi_b T_b -(i\theta^{a}[T_{a},T_{b}])\phi^{b},\nonumber \\
& =&\phi -(i\theta^{a}if_{abc}T_c)\phi^{b},\nonumber \\
& =&\phi -i\theta^{a}(if_{abc})T_c\phi^{b}.
\label{matrep}
\end{eqnarray}
Thus, Eqs.~(\ref{vecrep}),~(\ref{matrep}) with Eq.~(\ref{c}),
show that the both representations are equivalent.
In this thesis, we rely on the matrix representation (notation).

\chapter{Appendix C}

\section{Coupling constant calculation}
\begin{figure}
\begin{center}
\includegraphics[scale=0.4]{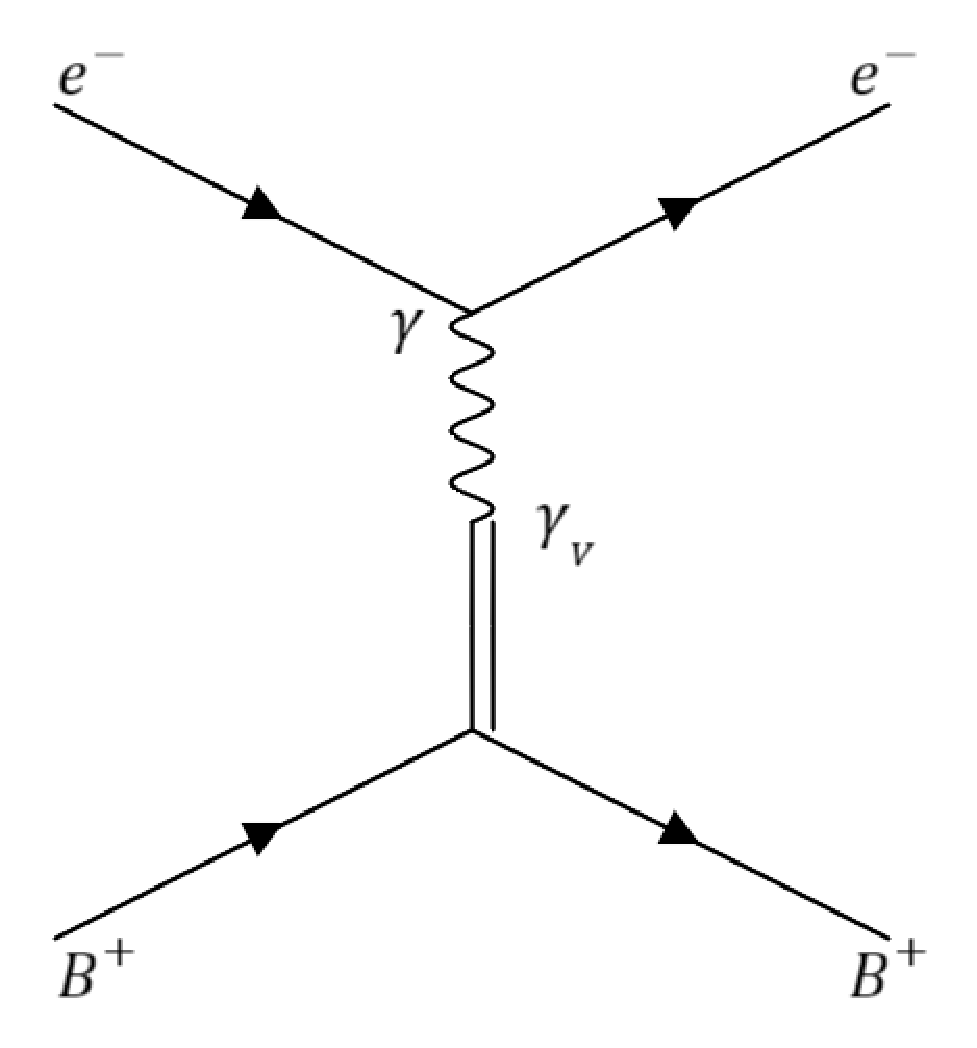}
\caption{ VDM feynman diagram
\label{vmd}}
\end{center}
\end{figure}
Developed by J.J. Sakurai in 1960s ~\cite{Sakurai:1960ju}, the VMD model describes interactions between energetic photons
and hadronic matter.

The coupling constants used in our calculation were obtained 
within the framework of the VMD model. 

The virtual photon in the process $e^-B^+\rightarrow e^-B^+$  
(and $e^-B^0\rightarrow e^-B^0$)
is coupled to vector mesons $\rho, \omega$, and $\Upsilon$, which are
then coupled to the bottom meson.

 The bellow equation is valid at zero momentum transfer,

\begin{eqnarray}
\sum_{V=\rho,\omega,\Upsilon}
\frac{\gamma_V~ f_{V B^+ B^+}}{m_V^2}=e. 
\end{eqnarray}

Where $\gamma_V$ is the photon-vector-meson mixing amplitude and
can be determined from the vector meson partial decay width to 
$e^+ e^-$, for example,
\begin{eqnarray}
\Gamma_{Vee}=\frac{\alpha\gamma_V^2}{3 m_V^3}.
\label{decay}
\end{eqnarray} 

Where $\alpha$ is the fine structure constant and its value is $\alpha=e^2/4\pi$.

Let us consider $f_{V B B} \equiv f_{V B^+ B^+}$ and thus
$f_{\rho B^+ B^+}=-f_{\rho B^+ B^+}$ 
from isospin symmetry, we then have:

\begin{eqnarray}
\frac{\gamma_\rho f_{\rho B B}}{m_\rho^2}=\frac {1}{2}e,\\
\frac{\gamma_\omega f_{\omega B B}}{m_\omega^2}=\frac {1}{6}e,\\
\frac{\gamma_\Upsilon f_{\Upsilon B B}}{m_\Upsilon^2}=\frac {1}{3}e.
\end{eqnarray} 

We can then, write:
\begin{eqnarray}
\gamma_{V}= m_{V} \sqrt{\frac{3m_{V} \Gamma_{Vee}}{\alpha}}
\end{eqnarray} 

Using the definitions of Eq.~(\ref{decay}) with $|e|=\sqrt{4\pi\alpha}$ in the above equations:

\begin{eqnarray}
f_{\Upsilon B B}& =&\frac {1}{3}\alpha \sqrt{\frac{4\pi m_\Upsilon}{3\Gamma_{(\Upsilon \ra e^+e^-)}}} \nonumber \\ 
& =&\frac{1}{3}\frac{1}{137.036}\sqrt{\frac{4\pi\times9460.30\times10^{3}}{3\times1.340}} \nonumber\\ 
& =&\frac{1}{3}(39.683474...) \nonumber \\
& =&13.2
\label{cc}
\end{eqnarray} 

$\Gamma_Ve^{+}e^{-}$ is experimental value obtained from PDG~\cite{PDG}. 
It is possible to obtain the coupling constants:

\begin{eqnarray}
f_{\rho B B}=2.48,\\
f_{\omega B B}=2.84,\\
f_{\Upsilon B B}=13.2.
\label{cc}
\end{eqnarray} 

Note that the value in Eq.~(\ref{eq36}) is the same as in Eq.~(\ref{cc}).






\noindent

\end{document}